\documentclass[12pt]{article}
\usepackage{amssymb}
\usepackage{color}
\usepackage{indentfirst}
\usepackage{amsmath}
\usepackage{enumerate}
\usepackage{cite}

\newtheorem{theorem}{Theorem}
\newtheorem{lemma}{Lemma}

\newtheorem{example}{Example}

\def\qi#1 {\fbox {\footnote {\ }}\ \footnotetext { From Qi: {\color{red}#1}}}

\begin{document}
\title{Weight distributions of two classes of linear codes based on Gaussian period and Weil sums}
\author{Xina Zhang\\
College of Statistics and Data Science\\ Lanzhou University of Finance and Economics\\
 Lanzhou,  Gansu 730070,  P.R. China\\
Email: zhangxina11@163.com}
\date{}
\maketitle

\begin{abstract}
In this paper, based on the theory of defining sets, two classes of at most six-weight linear codes over $\mathbb{F}_p$ are constructed. The weight distributions of the linear codes are determined by means of Gaussian period and Weil sums. In some case, there is an almost optimal code with respect to Griesmer bound, which is also an optimal one according to the online code table. The linear codes can also be employed to get secret sharing schemes.


\end{abstract}

\section{Introduction}
In this study, $p$ is an odd prime and assume $q=p^e$ for a positive integer $e$. Let $\mathbb{F}_p$ and $\mathbb{F}_q$ denote the finite field with $p$ and $q$ elements, respectively. We denote by $Tr$ the absolute trace function \cite{LN} from $\mathbb{F}_q$ onto $\mathbb{F}_p$, and use $\mathbb{F}^{*}_q$ and $\mathbb{F}^{*}_p$ to denote the multiplicative group of $\mathbb{F}_q$ and $\mathbb{F}_p$. Obviously, $\mathbb{F}_{q}^{*}=\mathbb{F}_{q}\setminus\{0\}$, and $\mathbb{F}_{p}^{*}=\mathbb{F}_{p}\setminus\{0\}$.

An $[n,k,d]$ \emph{linear code} $\mathcal{C}$ over $\mathbb{F}_q$ is a $k$-dimensional subspace of
$\mathbb{F}_q^n$ with minimum Hamming distance $d$. Let $A_i$ be the number of codewords with Hamming weight $i$ in a code $\mathcal{C}$. The weight enumerator of $\mathcal{C}$ is defined by
$$1+A_1z+A_2z^2+\ldots+A_nz^n,$$
and the sequence $(1, A_1, \ldots, A_n)$ is called the \emph{weight distribution} of $\mathcal{C}$\cite{HP}. If $|\{1\leq i\leq n: A_i\neq 0\}|=t,$ then we say $\mathcal{C}$ a $t$-weight code. In coding theory, the weight distribution of linear codes is an interesting research topic, as it contains important information as to estimate the error correcting capability and the probability of error detection and correction with respect to some algorithms.

Let $\mathcal{C}$ be an $[n,k,d]$ code over $\mathbb{F}_{q}$ with $k\geqslant1$, then the well-known \emph{Griesmer bound}\cite{HP} is given by
$$n\geqslant\sum_{i=0}^{k-1}\lceil\frac{d}{q^{i}}\rceil.$$

An $[n,k,d]$ code is called \emph{optimal} if no $[n,k,d+1]$ code exists, and is called \emph{almost optimal} if the $[n,k,d+1]$ code is optimal\cite{HY}.

One of the constructions of linear codes is based on a proper selection of a subset of finite fields\cite{DCS07}. That is, let $D = \{ {d_1},{d_2}, \ldots ,{d_n}\} \subseteq {F_q}$. A linear code of length $n$ over $\mathbb{F}_p$ is defined as
\begin{equation*}
\mathcal{C}_D =\{(Tr(xd_1),Tr(xd_2), \ldots ,Tr(xd_n)):x \in \mathbb{F}_q\},
\end{equation*}
the set $D$ is called the \emph{defining set} of linear code $\mathcal{C}_D$.  This construction approach is generic in the sense that many classes of optimal linear codes could be produced by selecting the proper defining sets\cite{KD,HY1,TXF,JLF,SY}.

By means of the construction method mentioned above, Xiang et al.\cite{CCK} constructed a class of linear codes and presented their weight distributions, with the defining set
$D = \{x\in\mathbb{F}_q^* : Tr(x^{p + 1}-x) = 0  \},$  where $p$ is an odd prime, $q=p^m.$
In this paper, we generalize the construction of the defining set, and obtain two classes of linear codes with at most six weights, which include some almost optimal codes. And making use of Weil sums\cite{R,S,C} and Gaussian period, we will determine not only the parameters but also weight distributions of these codes.

\section{Main Results}\label{theorem}
In this section, we present the main results, including the construction, the parameters and the weight distribution of the linear code $\mathcal{C}_{D_i}$. The proofs will be given in the following section.

We begin this section by selecting defining sets
\begin{equation}
D_{i}=\{ x \in \mathbb{F}_q : \mathrm{Tr}(x^{p+1}-x) \in C_i^{(2,p)} \}, i=0,1.
\end{equation}
to construct linear codes
\begin{equation}
\mathcal{C}_{D_{i}}=\{ \mathbf{c}(a)=(\mathrm{Tr}(ax))_{x \in D_{i}} : a \in \mathbb{F}_q \}, i=0,1.
\end{equation}
where $C_0^{(2,p)}$ and $C_1^{(2,p)}$ are the cyclotomic classes of order $2$ in $\mathbb{F}_p^*$\cite{ST}, also denote the sets of all squares and non-squares in $\mathbb{F}_p^*$, respectively. Let $q=p^e$ satisfying $p$ be an odd prime. Throughout the paper, $\eta$ is the quadratic character over $\mathbb{F}_{p}^{*}$. Then the weights and weight distributions of the linear codes are studied by utilizing some results of Weil sums\cite{S,C} and Gaussian period.

The following Theorems \ref{weight1}-\ref{weight9} are the main results of this paper.

\begin{theorem}\label{weight1}
If $e$ is odd and $p\mid e$, then the weight distribution of the codes $\mathcal{C}_{D_i}$\,$(i=0,1)$ with the parameters $[\frac{p-1}{2}p^{e-1}+(-1)^i\frac{p-1}{2}G^{e-1}, e]$ is listed in table \ref{1}.
Obviously, the codes are at most $5$-weight. $G=\sqrt{\eta(-1)p}$.
\end{theorem}
\begin{table}[!htp]
\begin{center}
\caption{The weight distribution of $\mathcal{C}_{D_i}$\,($i=0,1$) when $2\nmid e,\,p\mid e$}\label{1}
\begin{tabular}{ll}
\hline\noalign{\smallskip}
Weight  &  Multiplicity   \\
\noalign{\smallskip}
\hline\noalign{\smallskip}
$0$  &  1 \\
$\frac{(p-1)^2}{2}p^{e-2}$  &  $ p^{e-2}-1 $    \\
$\frac{(p-1)^2}{2}p^{e-2}+(-1)^i\frac{p-1}{2}G^{e-1}$  &  $(p-1)p^{e-2}$     \\
$\frac{(p-1)^2}{2}p^{e-2}+(-1)^i\frac{p-1}{2}G^{e-1}+(-1)^i\frac{p-1}{2}G^{e-3}$  &  $\frac{p-1}{2}p^{e-2}-(-1)^i\frac{p-1}{2}G^{e-1}+(1-\eta(-1))\frac{(p-1)^2}{4}p^{e-2}$     \\
$\frac{(p-1)^2}{2}p^{e-2}+(-1)^i\frac{p-1}{2}G^{e-1}-(-1)^i\frac{p-1}{2}G^{e-3}$  &  $\frac{p-1}{2}p^{e-1}+(-1)^i\frac{p-1}{2}G^{e-1}+(1+\eta(-1))\frac{(p-1)^2}{4}p^{e-2}$     \\
$\frac{(p-1)^2}{2}p^{e-2}+(-1)^i\frac{p-1}{2}G^{e-1}-(-1)^i\frac{p+1}{2}G^{e+3}$  &
$\frac{(p-1)^2}{2}p^{e-2}$     \\
\noalign{\smallskip}
\hline
\end{tabular}
\end{center}
\end{table}

\begin{theorem}\label{weight2}
If $e$ is odd and $\big(\frac{e}{p}\big)=(-1)^i$, $i=0,1$, then the weight distribution of the codes $\mathcal{C}_{D_i}$\,$(i=0,1)$ with the parameters $[\frac{p-1}{2}p^{e-1}-(-1)^iG^{e-1}, e]$ is listed in table \ref{2}.
Obviously, the codes are at most $6$-weight, where $G=\sqrt{\eta(-1)p}$, and $(\frac{\cdot}{\cdot})$ is the Legendre symbol.
\end{theorem}
\begin{table}[!htp]
\begin{center}
\caption{The weight distribution of $\mathcal{C}_{D_i}$\,($i=0,1$) when $2\nmid e,\,\big(\frac{e}{p}\big)=(-1)^i$}\label{2}
\begin{tabular}{ll}
\hline\noalign{\smallskip}
Weight  &  Multiplicity   \\
\noalign{\smallskip}
\hline\noalign{\smallskip}
$0$  &  1 \\
$\frac{(p-1)^2}{2}p^{e-2}$  &  $ p^{e-2}-1+(-1)^i\eta(-1)(p-1)G^{e-3}$    \\
$\frac{(p-1)^2}{2}p^{e-2}-(-1)^iG^{e-1}$  &  $(p-1)p^{e-2}-(-1)^i\eta(-1)(p-1)G^{e-3}$     \\
$\frac{(p-1)^2}{2}p^{e-2}-(-1)^iG^{e-1}-(-1)^i\eta(-1)\frac{p-1}{2}G^{e-3}$  &  $\frac{(p-1)^2}{4}p^{e-2}-(-1)^i\eta(-1)\frac{(p-1)^2}{4}G^{e-3}$    \\
$\frac{(p-1)^2}{2}p^{e-2}-(-1)^iG^{e-1}+(-1)^i\eta(-1)\frac{p-1}{2}G^{e-3}$  &
$\frac{(p-1)(p+3)}{4}p^{e-2}+(-1)^i\eta(-1)\frac{3(p-1)^2}{4}G^{e-3}$     \\
$\frac{(p-1)^2}{2}p^{e-2}-(-1)^iG^{e-1}+(-1)^i\eta(-1)\frac{p+1}{2}G^{e-3}$  &  $\frac{p^2-1}{4}p^{e-2}-(-1)^i\eta(-1)\frac{p^2-1}{4}G^{e-3}$     \\
$\frac{(p-1)^2}{2}p^{e-2}-(-1)^iG^{e-1}-(-1)^i\eta(-1)\frac{p+1}{2}G^{e-3}$  &
$\frac{(p-1)(p-3)}{4}p^{e-2}-(-1)^i\eta(-1)\frac{(p-1)(p-3)}{4}G^{e-3}$     \\
\noalign{\smallskip}
\hline
\end{tabular}
\end{center}
\end{table}

\begin{theorem}\label{weight3}
If $e$ is odd and $\big(\frac{e}{p}\big)=(-1)^{1-i}$, $i=0,1$, then the weight distribution of the codes $\mathcal{C}_{D_i}$\,$(i=0,1)$ with the parameters $[\frac{p-1}{2}p^{e-1}, e]$ is listed in table \ref{3}, where $(\frac{\cdot}{\cdot})$ is the Legendre symbol.
Obviously, the codes are at most $5$-weight.
\end{theorem}
\begin{table}[!htp]
\begin{center}
\caption{The weight distribution of $\mathcal{C}_{D_i}$\,($i=0,1$) when $2\nmid e,\,\big(\frac{e}{p}\big)=(-1)^{1-i}$}\label{3}
\begin{tabular}{ll}
\hline\noalign{\smallskip}
Weight  &  Multiplicity   \\
\noalign{\smallskip}
\hline\noalign{\smallskip}
$0$  &  1 \\
$\frac{(p-1)^2}{2}p^{e-2}$  &  $ p^{e-1}-1 $    \\
$\frac{(p-1)^2}{2}p^{e-2}+(-1)^i\eta(-1)\frac{p-1}{2}G^{e-3}$  &  $\frac{p^2-1}{4}p^{e-2}+(-1)^i\eta(-1)\frac{p^2-1}{4}G^{e-3}$     \\
$\frac{(p-1)^2}{2}p^{e-2}-(-1)^i\eta(-1)\frac{p-1}{2}G^{e-3}$  &  $\frac{p^2-1}{4}p^{e-2}-(-1)^i\eta(-1)\frac{(p-1)(3p-1)}{4}G^{e-3}$     \\
$\frac{(p-1)^2}{2}p^{e-2}-(-1)^i\frac{p+1}{2}G^{e-3}$  &  $\frac{(p-1)^2}{4}p^{e-2}+(-1)^i\eta(-1)\frac{(p-1)^2}{4}G^{e-3}$     \\
$\frac{(p-1)^2}{2}p^{e-2}+(-1)^i\frac{p+1}{2}G^{e-3}$  &  $\frac{(p-1)^2}{4}p^{e-2}+(-1)^i\eta(-1)\frac{(p-1)^2}{4}G^{e-3}$     \\
\noalign{\smallskip}
\hline
\end{tabular}
\end{center}
\end{table}

\begin{theorem}\label{weight4}
If $e$ is even, $e\equiv 2 \bmod 4$, and $p\mid e$, then the weight distribution of the codes $\mathcal{C}_{D_i}$\,$(i=0,1)$ with the parameters $[\frac{p-1}{2}p^{e-1}+\frac{p-1}{2}p^{\frac{e}{2}-1}, e]$ is listed in table \ref{4}.
Obviously, the codes are at most $4$-weight.
\end{theorem}
\begin{table}[!htp]
\begin{center}
\caption{The weight distribution of $\mathcal{C}_{D_i}$\,($i=0,1$) when $2\mid e,\,e\equiv 2 \bmod 4,\,p\mid e$}\label{4}
\begin{tabular}{ll}
\hline\noalign{\smallskip}
Weight  &  Multiplicity   \\
\noalign{\smallskip}
\hline\noalign{\smallskip}
$0$  &  1 \\
$\frac{(p-1)^2}{2}p^{e-2}$  &  $ \frac{p+1}{2}p^{e-2}-1-\frac{p-1}{2}p^{\frac{e}{2}-1} $    \\
$\frac{(p-1)^2}{2}p^{e-2}+\frac{p-1}{2}p^{\frac{e}{2}-1}$  &  $\frac{p^2-1}{2}p^{e-2}$     \\
$\frac{(p-1)^2}{2}p^{e-2}+(p-1)p^{\frac{e}{2}-1}$  &  $\frac{p-1}{2}p^{e-2}+\frac{p-1}{2}p^{\frac{e}{2}-1}$     \\
$\frac{(p-1)^2}{2}p^{e-2}+\frac{p-3}{2}p^{\frac{e}{2}-1}$  &  $\frac{(p-1)^2}{2}p^{e-2}$     \\
\noalign{\smallskip}
\hline
\end{tabular}
\end{center}
\end{table}

\begin{theorem}\label{weight5}
If $e$ is even, $e\equiv 2 \bmod 4$, and $\big(\frac{e}{p}\big)=(-1)^i$, $i=0,1$, then the weight distribution of the codes $\mathcal{C}_{D_i}$\,$(i=0,1)$ with the parameters $[\frac{p-1}{2}p^{e-1}-\frac{1+\eta(-1)p}{2}p^{\frac{e}{2}-1}, e]$ is listed in the case of $p\equiv 1 \bmod 4$ and $p\equiv 3 \bmod 4$ in table \ref{0} and \ref{6}, where $(\frac{\cdot}{\cdot})$ is the Legendre symbol.
Obviously, the codes are at most $5$-weight.
\end{theorem}
\begin{table}[!htp]
\begin{center}
\caption{The weight distribution of $\mathcal{C}_{D_i}$\,($i=0,1$) when $e\equiv 2 \bmod 4,\,\big(\frac{-e}{p}\big)=(-1)^i$}\label{0}
\begin{tabular}{ll}
\hline\noalign{\smallskip}
Weight  &  Multiplicity   \\
\noalign{\smallskip}
\hline\noalign{\smallskip}
$0$  &  1 \\
$\frac{(p-1)^2}{2}p^{e-2}$  &  $ p^{e-2}-1 $    \\
$\frac{(p-1)^2}{2}p^{e-2}-\frac{p+1}{2}p^{\frac{e}{2}-1}$  &  $\frac{p^2-1}{2}p^{e-2}-(p-1)p^{\frac{e}{2}-1}$     \\
$\frac{(p-1)^2}{2}p^{e-2}-\frac{p+3}{2}p^{\frac{e}{2}-1}$  &  $\frac{(p-1)(p-3)}{4}p^{e-2}-\frac{(p-1)(p-3)}{4}p^{\frac{e}{2}-1}$     \\
$\frac{(p-1)^2}{2}p^{e-2}-\frac{p-1}{2}p^{\frac{e}{2}-1}$  &  $\frac{p^2-1}{4}p^{e-2}+\frac{p^2-1}{4}p^{\frac{e}{2}-1}$     \\
$\frac{(p-1)^2}{2}p^{e-2}-p^{\frac{e}{2}-1}$  &  $(p-1)p^{e-2}$     \\
\noalign{\smallskip}
\hline
\end{tabular}
\end{center}
\end{table}

\begin{theorem}\label{weight6}
If $e$ is even, $e\equiv 2 \bmod 4$, and $\big(\frac{e}{p}\big)=(-1)^{1-i}$, $i=0,1$, then the weight distribution of the codes $\mathcal{C}_{D_i}$\,$(i=0,1)$ with the parameters $[\frac{p-1}{2}p^{e-1}-\frac{1-\eta(-1)p}{2}p^{\frac{e}{2}-1}, e]$ is listed in the case of $p\equiv 1 \bmod 4$ and $p\equiv 3 \bmod 4$ in table \ref{6} and \ref{0}, where $(\frac{\cdot}{\cdot})$ is the Legendre symbol.
Obviously, the codes are at most $5$-weight.
\end{theorem}
\begin{table}[!htp]
\begin{center}
\caption{The weight distribution of $\mathcal{C}_{D_i}$\,($i=0,1$) when $e\equiv 2 \bmod 4,\,\big(\frac{-e}{p}\big)=(-1)^{1-i}$}\label{6}
\begin{tabular}{ll}
\hline\noalign{\smallskip}
Weight  &  Multiplicity   \\
\noalign{\smallskip}
\hline\noalign{\smallskip}
$0$  &  1 \\
$\frac{(p-1)^2}{2}p^{e-2}$  &  $ p^{e-1}-1 $    \\
$\frac{(p-1)^2}{2}p^{e-2}+\frac{p-1}{2}p^{\frac{e}{2}-1}$  &  $\frac{p^2-1}{2}p^{e-2}$     \\
$\frac{(p-1)^2}{2}p^{e-2}+\frac{p-3}{2}p^{\frac{e}{2}-1}$  &  $\frac{(p-1)^2}{4}p^{e-2}-\frac{(p-1)^2}{4}p^{\frac{e}{2}-1}$     \\
$\frac{(p-1)^2}{2}p^{e-2}+\frac{p+1}{2}p^{\frac{e}{2}-1}$  &  $\frac{(p-1)^2}{4}p^{e-2}+\frac{(p-1)^2}{4}p^{\frac{e}{2}-1}$     \\
\noalign{\smallskip}
\hline
\end{tabular}
\end{center}
\end{table}

\begin{theorem}\label{weight7}
If $e$ is even, $e\equiv 0 \bmod 4$, and $p\mid e$, then the weight distribution of the codes $\mathcal{C}_{D_i}$\,$(i=0,1)$ with the parameters $[\frac{p-1}{2}p^{e-1}+\frac{p-1}{2}p^{\frac{e}{2}}, e]$ is listed in table \ref{7}.
Obviously, the codes are at most $5$-weight.
\end{theorem}
\begin{table}[!htp]
\begin{center}
\caption{The weight distribution of $\mathcal{C}_{D_i}$\,($i=0,1$) when $e\equiv 0 \bmod 4,\,p\mid e$}\label{7}
\begin{tabular}{ll}
\hline\noalign{\smallskip}
Weight  &  Multiplicity   \\
\noalign{\smallskip}
\hline\noalign{\smallskip}
$0$  &  1 \\
$\frac{(p-1)^2}{2}p^{e-2}+\frac{(p-1)^2}{2}p^{\frac{e}{2}-1}$  &  $p^e-p^{e-2}$    \\
$\frac{(p-1)^2}{2}p^{e-2}$  &  $ \frac{p+1}{2}p^{e-4}-1-\frac{p-1}{2}p^{\frac{e}{2}-2} $    \\
$\frac{(p-1)^2}{2}p^{e-2}+\frac{p-1}{2}p^{\frac{e}{2}}$  &  $\frac{p^2-1}{2}p^{e-4}$     \\
$\frac{(p-1)^2}{2}p^{e-2}+(p-1)p^{\frac{e}{2}}$  &  $\frac{p-1}{2}p^{e-4}+\frac{p-1}{2}p^{\frac{e}{2}-2}$     \\
$\frac{(p-1)^2}{2}p^{e-2}+\frac{p-3}{2}p^{\frac{e}{2}}$  &  $\frac{(p-1)^2}{2}p^{e-4}$     \\
\noalign{\smallskip}
\hline
\end{tabular}
\end{center}
\end{table}

\begin{theorem}\label{weight8}
If $e$ is even, $e\equiv 0 \bmod 4$, and $\big(\frac{e}{p}\big)=(-1)^i$, $i=0,1$, then the weight distribution of the codes $\mathcal{C}_{D_i}$\,$(i=0,1)$ with the parameters $[\frac{p-1}{2}p^{e-1}-\frac{1+\eta(-1)p}{2}p^{\frac{e}{2}}, e]$ is listed in table \ref{8} and \ref{9}, where $(\frac{\cdot}{\cdot})$ is the Legendre symbol.
Obviously, the codes are at most $6$-weight.
\end{theorem}
\begin{table}[!htp]
\begin{center}
\caption{The weight distribution of $\mathcal{C}_{D_i}$\,($i=0,1$) when $e\equiv 0 \bmod 4,\,\big(\frac{-e}{p}\big)=(-1)^i$}\label{8}
\begin{tabular}{ll}
\hline\noalign{\smallskip}
Weight  &  Multiplicity   \\
\noalign{\smallskip}
\hline\noalign{\smallskip}
$0$  &  1 \\
$\frac{(p-1)^2}{2}p^{e-2}-\frac{p^2-1}{2}p^{\frac{e}{2}-1}$  &  $p^e-p^{e-2}$   \\
$\frac{(p-1)^2}{2}p^{e-2}$  &  $ p^{e-4}-1 $    \\
$\frac{(p-1)^2}{2}p^{e-2}-\frac{p+1}{2}p^{\frac{e}{2}}$  &  $\frac{p^2-1}{2}p^{e-4}-(p-1)p^{\frac{e}{2}-2}$     \\
$\frac{(p-1)^2}{2}p^{e-2}-\frac{p+3}{2}p^{\frac{e}{2}}$  &  $\frac{(p-1)(p-3)}{4}p^{e-4}-\frac{(p-1)(p-3)}{4}p^{\frac{e}{2}-2}$     \\
$\frac{(p-1)^2}{2}p^{e-2}-\frac{p-1}{2}p^{\frac{e}{2}}$  &  $\frac{p^2-1}{4}p^{e-4}+\frac{p^2-1}{4}p^{\frac{e}{2}-2}$     \\
$\frac{(p-1)^2}{2}p^{e-2}-p^{\frac{e}{2}}$  &  $(p-1)p^{e-4}$     \\
\noalign{\smallskip}
\hline
\end{tabular}
\end{center}
\end{table}

\begin{theorem}\label{weight9}
If $e$ is even, $e\equiv 0 \bmod 4$, and $\big(\frac{e}{p}\big)=(-1)^{1-i}$, $i=0,1$, then the weight distribution of the codes $\mathcal{C}_{D_i}$\,$(i=0,1)$ with the parameters $[\frac{p-1}{2}p^{e-1}-\frac{1-\eta(-1)p}{2}p^{\frac{e}{2}}, e]$ is listed in table \ref{9} and \ref{8}, where $(\frac{\cdot}{\cdot})$ is the Legendre symbol.
Obviously, the codes are at most $6$-weight.
\end{theorem}
\begin{table}[!htp]
\begin{center}
\caption{The weight distribution of $\mathcal{C}_{D_i}$\,($i=0,1$) when $e\equiv 0 \bmod 4,\,\big(\frac{-e}{p}\big)=(-1)^{1-i}$}\label{9}
\begin{tabular}{ll}
\hline\noalign{\smallskip}
Weight  &  Multiplicity   \\
\noalign{\smallskip}
\hline\noalign{\smallskip}
$0$  &  1 \\
$\frac{(p-1)^2}{2}p^{e-2}+\frac{(p-1)^2}{2}p^{\frac{e}{2}-1}$  &  $p^e-p^{e-2}$    \\
$\frac{(p-1)^2}{2}p^{e-2}$  &  $ p^{e-3}-1 $    \\
$\frac{(p-1)^2}{2}p^{e-2}+\frac{p-1}{2}p^{\frac{e}{2}}$  &  $\frac{p^2-1}{2}p^{e-4}$     \\
$\frac{(p-1)^2}{2}p^{e-2}+\frac{p-3}{2}p^{\frac{e}{2}}$  &  $\frac{(p-1)^2}{4}p^{e-4}-\frac{(p-1)^2}{4}p^{\frac{e}{2}-2}$     \\
$\frac{(p-1)^2}{2}p^{e-2}+\frac{p+1}{2}p^{\frac{e}{2}}$  &  $\frac{(p-1)^2}{4}p^{e-4}+\frac{(p-1)^2}{4}p^{\frac{e}{2}-2}$     \\
\noalign{\smallskip}
\hline
\end{tabular}
\end{center}
\end{table}

The followings are some examples about our results verified by Magma.

\begin{example}\label{example1}
If $(p,e)=(3, 3)$, then by Theorem \ref{weight1}, the code $\mathcal{C}_{D_0}$ has parameters $[6,3,3]$ with weight enumerator $1+6z^{3}+12z^{4}+6z^{5}+2z^{6}$, the code $\mathcal{C}_{D_1}$ has parameters $[12,3,6]$ with weight enumerator $1+2z^{6}+6z^{7}+6z^{8}+6z^{9}+6z^{10}$,
which confirmed the result by Magma. According to Griesmer bound, also with respect to the code table\cite{YZ}, the code $\mathcal{C}_{D_{20}}$ is optimal.
\end{example}

\begin{example}\label{example2}
If $(p,e)=(5,2)$, then by Theorem \ref{weight6}, the code $\mathcal{C}_{D_0}$ has parameters $[12,2,8]$ with weight enumerator
$1+4z^{8}+12z^{10}+8z^{11}$, the code $\mathcal{C}_{D_1}$ has parameters $[7,2,5]$ with weight enumerator
$1+8z^{5}+12z^{6}+4z^{7}$,
which confirmed the result by Magma. According to Griesmer bound, also with respect to the code table\cite{YZ}, the code $\mathcal{C}_{D_{21}}$ is optimal.
\end{example}

\begin{example}\label{example3}
If $(p,e)=(7, 4)$, then by Theorem \ref{weight9}, the code $\mathcal{C}_{D_0}$\,$(i=0,1)$ has parameters $[1176,4,882]$ with weight enumerator $1+6z^{882}+2352z^{1008}+24z^{1029}+18z^{1078}$, the code $\mathcal{C}_{D_1}$\,$(i=0,1)$ has parameters $[833,4,686]$ with weight enumerator $1+18z^{686}+2352z^{714}+24z^{735}+6z^{833}$,
which confirmed the result by Magma.
\end{example}

\section{Preliminaries and Auxiliary lemmas}\label{preliminaries}
In this section, we present some facts on exponential sums, that will be needed in calculating the weight enumerator of the codes defined in this article.

An additive character of $\mathbb{F}_q$ is a non-zero function $\chi$ from $\mathbb{F}_q$ to the set of complex numbers of absolute value $1$ such that
$\chi(x+y)=\chi(x)\chi(y)$ for any pair $(x,y) \in \mathbb{F}_q^2$. For each $u \in \mathbb{F}_q$, the function
$$\chi_u(v)=\zeta_{p}^{Tr(uv)},~v \in \mathbb{F}_q$$
denotes an additive character of $\mathbb{F}_q$, where $\zeta_{p}=e^{2\pi i/p}$ is a primitive $p$-th root of unity and $i=\sqrt{-1}$.  Since $\chi_0(v)=1$ for all $v \in \mathbb{F}_q$, which is the trivial additive character of $\mathbb{F}_q$. We call $\chi_1$ the canonical additive character of $\mathbb{F}_q$ and we have $\chi_u(x)=\chi_1(ux)$ for all $u\in\mathbb{F}_q$. The additive character satisfies the orthogonal property \cite{LN}, that is
\begin{eqnarray*}
\sum_{v \in \mathbb{F}_q} \chi_u(v)=\left\{
\begin{array}{ll}
q,   & u=0,\\  
0,    & u\neq0.\\  
\end{array}
\right.
\end{eqnarray*}

Let $h$ be a fixed primitive element of $\mathbb{F}_q$. For each $j=0, 1, \ldots, q-2,$ the function $\lambda_j(h^k)=e^{2\pi ijk/(q-1)}$ for $k=0, 1, \ldots, q-2$ defines a multiplicative character of $\mathbb{F}_q$, we extend these characters by setting $\lambda_j(0)=0$. Let $q$ be odd. For $j=(q-1)/2$ and $v \in \mathbb{F}_{q}^{*}$, we have
\begin{equation*}
  \lambda_{(q-1)/2}(v) =
  \begin{cases}
    1, & \text{if $v$ is the square of an element of $\mathbb{F}_{q}^{*}$,}  \\
    -1, & \text{otherwise,}
  \end{cases}
\end{equation*}
which is called the quadratic character of $\mathbb{F}_{q}$, and is denoted by $\eta'$ in the sequel.
We call $\eta'=\lambda_{(q-1)/2}$ and $\eta=\lambda_{(p-1)/2}$ are the quadratic characters over $\mathbb{F}_q$ and $\mathbb{F}_p$, respectively. The quadratic Gauss sums over $\mathbb{F}_q$ and $\mathbb{F}_p$ are defined respectively by
$$G'(\eta')=\sum\limits_{v \in \mathbb{F}_q}\eta'(v)\chi'_1(v) \quad \mathrm{and} \quad G(\eta)=\sum\limits_{v \in \mathbb{F}_p}\eta(v)\chi_1(v),$$
where $\eta$ and $\chi_1$ are the canonical multiplicative and additive characters of $\mathbb{F}_p$, respectively. Moreover, it is well known that $G'=(-1)^{e-1}\sqrt{p^{*}}^e$ and $G=\sqrt{p^{*}}$, where $p^{*}=\eta(-1)p.$

The following are some basic facts on exponential sums.

\begin{lemma}\label{quadratic sums}(\cite{LN}, Theorem 5.33)
If $f(x)=a_{2}x^{2}+a_{1}x+a_{0} \in \mathbb{F}_{q}[x],$ where $a_{2}\neq 0,$ then
$$\sum_{x \in \mathbb{F}_{q}}\zeta_{p}^{Tr(f(x))}=\zeta_{p}^{Tr(a_{0}-a_{1}^{2}(4a_{2})^{-1})}\eta'(a_{2})G'(\eta'),$$
where $\eta'$ is the quadratic character of $\mathbb{F}_{q}$.
\end{lemma}

\begin{lemma}\label{quadratic character}(\cite{LN}, Theorem 5.48)
With the notation above, we have
\begin{equation*}
  \sum_{x \in \mathbb{F}_{q}}\eta'(f(x)) =
  \begin{cases}
    -\eta'(a_{2}), & a_{1}^{2}-4a_{0}a_{2}\neq 0,\\ 
    (q-1)\eta'(a_{2}), & a_{1}^{2}-4a_{0}a_{2}=0.  
  \end{cases}
\end{equation*}
\end{lemma}

For $\alpha, \beta \in \mathbb{F}_{q}$ and any positive integer $l$, the Weil sums       $S(\alpha, \beta)$ is defined by
$$S(\alpha, \beta)= \sum_{x \in \mathbb{F}_{q}}\zeta_{p}^{Tr(\alpha x^{p^{l}+1}+\beta x)}.$$

We will show some results of $S(\alpha, \beta)$ for $\alpha\neq 0$ and $q$ odd.

\begin{lemma}\label{weil sums5}(\cite{CCK})
Let $e$ be odd and $p$ be an odd prime. $f(x)=\alpha^px^{p^2}+\alpha x \in \mathbb{F}_q[x]$ and $\beta \in \mathbb{F}_q$, then $f(x)$ is a permutation polynomial over $\mathbb{F}_{q}$ with $q=p^e$, and
\begin{equation*}
S(\alpha,\beta)=G^e\eta'(\alpha)\zeta_p^{\mathrm{Tr}(-\alpha x_0^{p+1})},
\end{equation*}
where $G=G(\eta)=\sqrt{p^*}=\sqrt{\eta(-1)p}=\sqrt{(-1)^{\frac{p-1}{2}}p}$, $x_0$ is the unique solution of the equation $f(x)=-\beta^p$.
Especially,
\begin{equation*}
S(\alpha,0)=G^e\eta'(\alpha).
\end{equation*}
\end{lemma}

\begin{lemma}\label{weil sums1}(\cite{S}, Theorem 2)
Let $s=(l,e)$ and $e/s$ be even with $e=2m$. Then
\begin{equation*}
  S(\alpha, 0) =
  \begin{cases}
   (-1)^{m/s}p^m , & \alpha^{(q-1)/(p^s+1)} \neq (-1)^{m/s},\\   
    (-1)^{m/s+1}p^{m+s}, & \alpha^{(q-1)/(p^s+1)}=(-1)^{m/s}.   
  \end{cases}
\end{equation*}
\end{lemma}

\begin{lemma}\label{weil sums2}(\cite{C}, Theorem 4.7)
Let $\beta \neq 0$ and $e/s$ be even with $e=2m.$ Then $S(\alpha, \beta)=0$ unless the equation $\alpha^{p^{l}}X^{p^{2l}}+\alpha X=-\beta^{p^{l}}$ is solvable. There are two possibilities.
\begin{enumerate}
  \item If $\alpha^{(q-1)/(p^s+1)} \neq (-1)^{m/s},$ then for any choice of $\beta \in \mathbb{F}_q,$ the equation has a unique solution $x_{0}$ and $$S(\alpha, \beta)=(-1)^{m/s}p^{m}\zeta_{p}^{Tr(-\alpha x_{0}^{p^{l}+1})}$$
  \item If $\alpha^{(q-1)/(p^s+1)}=(-1)^{m/s}$ and if the equation is solvable with some solution $x_{0}$, then
  $$S(\alpha, \beta)=(-1)^{m/s+1}p^{m+s}\zeta_{p}^{Tr(-\alpha x_{0}^{p^{l}+1})}$$
\end{enumerate}
\end{lemma}

\begin{lemma}\label{weil sums3}(\cite{S}, Theorem 4.1)
For $e=2m$, the equation $\alpha^{p^{l}} X^{p^{2l}}+\alpha X=0$ is solvable for $X \in \mathbb{F}_{q}^{*}$ if and only if $e/s$ is even and $\alpha^{(q-1)/(p^{s}+1)}=(-1)^{m/s}$. In such cases, there are $p^{2s}-1$ non-zero solutions.
\end{lemma}

There is the fact that $\alpha^{p^{l}}X^{p^{2l}}+\alpha X$ is a permutation polynomial over $\mathbb{F}_{q}$ with $q=p^e$ if and only if $e/s$ is odd or $e/s$ is even with $e=2m$ and $\alpha^{(q-1)/(p^s+1)} \neq (-1)^{m/s}$.

\begin{lemma}\label{weil sums4}(\cite{QFKD})
Let $f(X)=X^{p^{2l}}+X$ and $$S=\{\beta \in \mathbb{F}_{q}: f(X)=-\beta^{p^{l}} is\;solvable\;in\;\mathbb{F}_{q}\}.$$
If $m/s \equiv 0 \bmod 2$, then $|S|=p^{e-2s}$.
\end{lemma}

\begin{lemma}\label{Gaussian period}(\cite{GM})
Let $p$ an odd prime. $\forall x \in \mathbb{F}_p$, the quadratic Gaussian period over $\mathbb{F}_p$ is given by
\begin{equation*}
  \rho_i^{(2,p)}=\sum\limits_{x \in C_i^{(2,p)}}\chi_1(x),i=0,1.
\end{equation*}
The value of the quadratic Gaussian period is
\begin{equation*}
\rho_0^{(2,p)}=\frac{-1+\sqrt{p^*}}{2}=
\begin{cases}
\frac{-1+\sqrt{p}}{2}, & \text{if $p \equiv 1 \bmod 4$},\\
\frac{-1+\sqrt{-p}}{2}, & \text{if $p \equiv 3 \bmod 4$},
\end{cases}
\end{equation*}
and
\begin{equation*}
\rho_1^{(2,p)}=-1-\rho_0^{(2,p)}.
\end{equation*}
Obviously,
\begin{equation*}
2\rho_0^{(2,p)}+1=G(\eta)=G.
\end{equation*}
\end{lemma}

\begin{lemma}\label{change}(\cite{KD})
Let $p$ an odd prime, $q=p^e$. $\eta'$ and $\eta$ is the quadratic character over $\mathbb{F}_q^*$ and $\mathbb{F}_p^*$, respectively. $\forall y \in \mathbb{F}_p^*$, then
\begin{equation*}
\eta'(y)=
\begin{cases}
\eta(y), &  2\nmid e,\\
1, &  2\mid e.
\end{cases}
\end{equation*}
\end{lemma}

\begin{lemma}\label{number}(\cite{BRK})
Let $p$ be an odd prime and $c$ be an integer not divisible by $p$. Let $\epsilon_1=\pm 1$ and $\epsilon_2=\pm 1$. Then the number of integers $n$ such that $0\leq n \leq p-1$ and
\begin{equation*}
\big(\frac{n}{p}\big)=\epsilon_1,~\big(\frac{n+c}{p}\big)=\epsilon_2
\end{equation*}
is
\begin{equation*}
\frac{1}{4}\left[p-2-\epsilon_1\big(\frac{-c}{p}\big)-\epsilon_2\big(\frac{c}{p}\big)-\epsilon_1\epsilon_2\right].
\end{equation*}
\end{lemma}

\section{The proofs of the main results}\label{proofs}

For $i=0,1$, the lengths of $\mathcal{C}_{D_i}$\,$(i=0,1)$ are $n_i=|D_i|$, then
\begin{eqnarray}
n_i&=&|\{ x \in \mathbb{F}_q : \mathrm{Tr}(x^{p+1}-x) \in C_i^{(2,p)} \}|\nonumber\\
&=&\sum_{c \in C_{i}^{(2,p)}}|\{ x \in \mathbb{F}_q : \mathrm{Tr}(x^{p+1}-x)=c \}|\nonumber\\
&=&\sum_{c \in C_{i}^{(2,p)}}\sum_{x \in \mathbb{F}_q}\frac{1}{p}\sum_{y \in \mathbb{F}_p}\zeta_p^{y(\mathrm{Tr}(x^{p+1}-x)-c)}\nonumber\\
&=&\frac{1}{p}\sum_{c \in C_{i}^{(2,p)}}\sum_{x \in \mathbb{F}_q}(1+\sum_{y \in \mathbb{F}_p^*}\zeta_p^{y(\mathrm{Tr}(x^{p+1}-x)-c)})\nonumber\\
&=&\frac{p-1}{2}p^{e-1}+\frac{1}{p}\delta_{1,i},\label{5.2.1}
\end{eqnarray}
where
\begin{equation}\label{5.2.2}
\delta_{1,i}=\sum_{c \in C_{i}^{(2,p)}}\sum_{y \in \mathbb{F}_p^*}\zeta_p^{-cy}\sum_{x \in \mathbb{F}_q}\zeta_p^{\mathrm{Tr}(yx^{p+1}-yx)}.
\end{equation}

For any $a \in \mathbb{F}_q$ and any codeword $\mathbf{c}(a)\in \mathcal{C}_{D_i}$, to determine the weight enumerators of $\mathcal{C}_{D_i}$\,$(i=0,1)$, let
\begin{equation}\label{5.2.3}
T_i=|\{ x \in \mathbb{F}_q : \mathrm{Tr}(x^{p+1}-x) \in C_i^{(2,p)}, \mathrm{Tr}(ax)=0 \}|
\end{equation}
then we can deduce
\begin{equation}\label{5.2.4}
wt_i(\mathbf{c}(a))=n_i-T_i,
\end{equation}
where $wt_i(\mathbf{c}(a))$($i=0,1$) is Hamming weight of the codeword $\mathbf{c}(a)$.

For $a \in \mathbb{F}_q^*$, by the orthogonal property of additive character, we have
\begin{eqnarray}
T_i&=&\sum_{c \in C_i^{(2,p)}}\sum_{x \in \mathbb{F}_q}(\frac{1}{p}\sum_{y \in \mathbb{F}_p}\zeta_p^{y(\mathrm{Tr}(x^{p+1}-x)-c)})(\frac{1}{p}\sum_{z \in \mathbb{F}_p}\zeta_p^{z\mathrm{Tr}(ax)})\nonumber\\
&=&\frac{1}{p^2}\sum_{c \in C_i^{(2,p)}}\sum_{x \in \mathbb{F}_q}(1+\sum_{y \in \mathbb{F}_p^*}\zeta_p^{y(\mathrm{Tr}(x^{p+1}-x)-c)})(1+\sum_{z \in \mathbb{F}_p^*}\zeta_p^{z\mathrm{Tr}(ax)})\nonumber\\
&=&\frac{p-1}{2}p^{e-2}+\frac{1}{p^2}(\delta_{1,i}+\delta_{2,i}+\delta_{3,i}),\label{5.2.5}
\end{eqnarray}
where
\begin{eqnarray}
\delta_{2,i}&=&\sum_{c \in C_i^{(2,p)}}\sum_{z \in \mathbb{F}_p^*}\sum_{x \in \mathbb{F}_q}\zeta_p^{\mathrm{Tr}(azx)},\nonumber\\
\delta_{3,i}&=&\sum_{c \in C_i^{(2,p)}}\sum_{y \in \mathbb{F}_p^*}\zeta_p^{-cy}\sum_{z \in \mathbb{F}_p^*}\sum_{x \in \mathbb{F}_q}\zeta_p^{\mathrm{Tr}(yx^{p+1}+(az-y)x)}.\label{5.2.6}
\end{eqnarray}

By the orthogonal property of additive character, we have $\delta_{2,i}=0$.

Then
\begin{equation}\label{5.2.7}
wt_i(\mathbf{c}(a))=\frac{(p-1)^2}{2}p^{e-2}+\frac{1}{p}\delta_{1,i}-\frac{1}{p^2}(\delta_{1,i}+\delta_{3,i}).
\end{equation}

The following Lemmas are essential to determine the lengths                                                                                                                                                                                                                                                                                                                                                                                                                                                             and weight distributions of $\mathcal{C}_{D_i}$\,$(i=0,1)$.

\begin{lemma}\label{first}
For $i=0,1$,
\begin{eqnarray}\label{5.2.8}
\delta_{1,i}&=& \left\{
\begin{array}{ll}
(-1)^i\eta(-1)\frac{p-1}{2}G^{e+1}, & 2\nmid e,\,p\mid e,  \\
    -\frac{(-1)^{i}+\eta(e)}{2}\eta(-1)G^{e+1}, & 2\nmid e,\,p\nmid e, \\
    \frac{p-1}{2}p^{\frac{e}{2}}, & 2\mid e,\,e \equiv 2 \bmod 4,\,p\mid e,  \\
    -\frac{(-1)^i\eta(-e)p+1}{2}p^{\frac{e}{2}}, & 2\mid e,\,e \equiv 2 \bmod 4,\,p\nmid e, \\
    \frac{p-1}{2}p^{\frac{e}{2}+1}, & 2\mid e,\,e \equiv 0 \bmod 4,\,p\mid e,  \\
    -\frac{(-1)^i\eta(-e)p+1}{2}p^{\frac{e}{2}+1}, & 2\mid e,\,e \equiv 0 \bmod 4,\,p\nmid e.\\
\end{array}
\right.
\end{eqnarray}
\end{lemma}
\textbf{Proof:} With Weil sums, we have
\begin{eqnarray}
\delta_{1,i}&=&\sum_{c \in C_{i}^{(2,p)}}\sum_{y \in \mathbb{F}_p^*}\zeta_p^{-cy}\sum_{x \in \mathbb{F}_q}\zeta_p^{\mathrm{Tr}(yx^{p+1}-yx)}\nonumber\\
&=&\sum_{c \in C_{i}^{(2,p)}}\sum_{y \in \mathbb{F}_p^*}\zeta_p^{-cy}S(y,-y).\label{5.2.9}
\end{eqnarray}
Note that $x=\frac{1}{2}$ is the unique solution of the equation $y^px^{p^2}+yx=-(-y)^p$ in $\mathbb{F}_q$. In fact, $y \in \mathbb{F}_p^*:\,y^p=y$, $(-1)^p=-1$, the equation $y^px^{p^2}+yx=-(-y)^p$ is equivalent to the equation $x^{p^2}+x=1$,
thus $\frac{1}{2} \in \mathbb{F}_p^*\subset\mathbb{F}_q^*$.

From Lemma \ref{weil sums5}, Lemma \ref{weil sums2} and Lemma \ref{change},
\begin{equation}\label{5.2.10}
\begin{aligned}
S(y,-y)&=
\begin{cases}
G^e\eta(y)\zeta_p^{-\frac{ey}{4}}, & ~\text{$2\nmid e$,}  \\
-p^{\frac{e}{2}}\zeta_p^{-\frac{ey}{4}}, & ~\text{$2\mid e,\,e\equiv 2 \bmod 4$,}  \\
-p^{\frac{e}{2}+1}\zeta_p^{-\frac{ey}{4}}, & ~\text{$2\mid e,\,e\equiv 0 \bmod 4$.}
\end{cases}
\end{aligned}
\end{equation}
Applying these values into Eq.(\ref{5.2.9}), we have
\begin{equation}\label{5.2.0}
\begin{aligned}
\delta_{1,i}&=
\begin{cases}
G^e\sum\limits_{y \in \mathbb{F}_p^*}\zeta_p^{-\frac{ey}{4}}\eta(y)\sum\limits_{c \in C_i^{(2,p)}}\zeta_p^{-cy}, & ~\text{$2\nmid e$,}  \\
-p^{\frac{e}{2}}\sum\limits_{y \in \mathbb{F}_p^*}\zeta_p^{-\frac{ey}{4}}\sum\limits_{c \in C_i^{(2,p)}}\zeta_p^{-cy}, & ~\text{$2\mid e,\,e\equiv 2 \bmod 4$,}  \\
-p^{\frac{e}{2}+1}\sum\limits_{y \in \mathbb{F}_p^*}\zeta_p^{-\frac{ey}{4}}\sum\limits_{c \in C_i^{(2,p)}}\zeta_p^{-cy}, & ~\text{$2\mid e,\,e\equiv 0 \bmod 4$.}
\end{cases}
\end{aligned}
\end{equation}
With the definition of Gaussian period, we have
\begin{eqnarray*}
\delta_{1,i}&=& \left\{
\begin{array}{ll}
G^e\eta(-1)(\sum\limits_{-y \in C_i^{(2,p)}}\eta(-y)\rho_0+\sum\limits_{-y \in C_{1-i}^{(2,p)}}\eta(-y)\rho_1), & 2\nmid e,\,p\mid e,  \\
G^e\eta(-1)(\sum\limits_{-y \in C_i^{(2,p)}}\zeta_p^{-\frac{ey}{4}}\eta(-y)\rho_0+\sum\limits_{-y \in C_{1-i}^{(2,p)}}\zeta_p^{-\frac{ey}{4}}\eta(-y)\rho_1), & 2\nmid e,\,p\nmid e,  \\
-p^{\frac{e}{2}}\sum\limits_{c \in C_i^{(2,p)}}(\sum\limits_{y \in \mathbb{F}_p^*}\zeta_p^{-cy}), & 2\mid e,\,e\equiv 2 \bmod 4,\,p\mid e,  \\
-p^{\frac{e}{2}}(\sum\limits_{-y \in C_i^{(2,p)}}\zeta_p^{-\frac{ey}{4}}\rho_0+\sum\limits_{-y \in C_{1-i}^{(2,p)}}\zeta_p^{-\frac{ey}{4}}\rho_1), & 2\mid e,\,e\equiv 2 \bmod 4,\,p\nmid e,  \\
-p^{\frac{e}{2}+1}\sum\limits_{c \in C_i^{(2,p)}}(\sum\limits_{y \in \mathbb{F}_p^*}\zeta_p^{-cy}), & 2\mid e,\,e\equiv 0 \bmod 4,\,p\mid e,  \\
-p^{\frac{e}{2}+1}(\sum\limits_{-y \in C_i^{(2,p)}}\zeta_p^{-\frac{ey}{4}}\rho_0+\sum\limits_{-y \in C_{1-i}^{(2,p)}}\zeta_p^{-\frac{ey}{4}}\rho_1), & 2\mid e,\,e\equiv 0 \bmod 4,\,p\nmid e.\\
\end{array}
\right.
\end{eqnarray*}
By the orthogonal property of additive character, we have
\begin{eqnarray*}
\delta_{1,i}&=& \left\{
\begin{array}{ll}
(-1)^i\eta(-1)G^e\frac{p-1}{2}(\rho_0-\rho_1), & 2\nmid e,\,p\mid e, \\
(-1)^i\eta(-1)G^e(\rho_0^2-\rho_1^2), & 2\nmid e,\,e \in C_i^{(2,p)},  \\
(-1)^i\eta(-1)G^e(\rho_1\rho_0-\rho_0\rho_1), & 2\nmid e,\,e \in C_{1-i}^{(2,p)},  \\
\frac{p-1}{2}p^{\frac{e}{2}}, & 2\mid e,\,e\equiv 2 \bmod 4,\,p\mid e,  \\
-p^{\frac{e}{2}}(\rho_0^2+\rho_1^2), & 2\mid e,\,e\equiv 2 \bmod 4,\,e \in C_i^{(2,p)},  \\
-p^{\frac{e}{2}}(2\rho_0\rho_1), & 2\mid e,\,e\equiv 2 \bmod 4,\,e \in C_{1-i}^{(2,p)},  \\
\frac{p-1}{2}p^{\frac{e}{2}+1}, & 2\mid e,\,e\equiv 0 \bmod 4,\,p\mid e, \\
-p^{\frac{e}{2}+1}(\rho_0^2+\rho_1^2), & 2\mid e,\,e\equiv 0 \bmod 4,\,e \in C_i^{(2,p)},  \\
-p^{\frac{e}{2}+1}(2\rho_0\rho_1), & 2\mid e,\,e\equiv 0 \bmod 4,\,e \in C_{1-i}^{(2,p)}.\\
\end{array}
\right.
\end{eqnarray*}
With Lemma \ref{Gaussian period},
\begin{eqnarray}\label{5.2.11}
\delta_{1,i}&=& \left\{
\begin{array}{ll}
    (-1)^i(\eta(-1))^{\frac{e+3}{2}}\frac{p-1}{2}p^{\frac{e+1}{2}}, & 2\nmid e,\,p\mid e, \\
    (-1)^{1+i}(\eta(-1))^{\frac{e+3}{2}}p^{\frac{e+1}{2}}, & 2\nmid e,\,e \in C_i^{(2,p)}, \\
    0, & ~\text{$2\nmid e,\,e \in C_{1-i}^{(2,p)}$,} \\
    \frac{p-1}{2}p^{\frac{e}{2}}, & 2\mid e,\,e \equiv 2 \bmod 4,\,p\mid e, \\
    -\frac{1+\eta(-1)p}{2}p^{\frac{e}{2}}, & 2\mid e,\,e \equiv 2 \bmod 4,\,e \in C_i^{(2,p)}, \\
    -\frac{1-\eta(-1)p}{2}p^{\frac{e}{2}}, & 2\mid e,\,e \equiv 2 \bmod 4,\,e \in C_{1-i}^{(2,p)}, \\
    \frac{p-1}{2}p^{\frac{e}{2}+1}, & 2\mid e,\,e \equiv 0 \bmod 4,\,p\mid e,  \\
    -\frac{1+\eta(-1)p}{2}p^{\frac{e}{2}+1}, & 2\mid e,\,e \equiv 0 \bmod 4,\,e \in C_i^{(2,p)}, \\
    -\frac{1-\eta(-1)p}{2}p^{\frac{e}{2}+1}, & 2\mid e,\,e \equiv 0 \bmod 4,\,e \in C_{1-i}^{(2,p)}.\\
\end{array}
\right.
\end{eqnarray}
After simplification, we can get the desired results.\hfill$\Box$

\begin{lemma}\label{length}
The length of the code $\mathcal{C}_{D_i}$\,$(i=0,1)$ is
\begin{eqnarray}\label{5.2.12}
n_i&=& \left\{
\begin{array}{ll}
\frac{p-1}{2}p^{e-1}+(-1)^i\frac{p-1}{2}G^{e-1}, & 2\nmid e,\,p\mid e,  \\
\frac{p-1}{2}p^{e-1}-\frac{(-1)^{i}+\eta(e)}{2}G^{e-1}, & 2\nmid e,\,p\nmid e, \\
\frac{p-1}{2}p^{e-1}+\frac{p-1}{2}p^{\frac{e}{2}-1}, & 2\mid e,\,e \equiv 2 \bmod 4,\,p\mid e,  \\
\frac{p-1}{2}p^{e-1}-\frac{(-1)^i\eta(-e)p+1}{2}p^{\frac{e}{2}-1}, & 2\mid e,\,e \equiv 2 \bmod 4,\,p\nmid e, \\
\frac{p-1}{2}p^{e-1}+\frac{p-1}{2}p^{\frac{e}{2}}, & 2\mid e,\,e \equiv 0 \bmod 4,\,p\mid e,  \\
\frac{p-1}{2}p^{e-1}-\frac{(-1)^i\eta(-e)p+1}{2}p^{\frac{e}{2}}, & 2\mid e,\,e \equiv 0 \bmod 4,\,p\nmid e.\\
\end{array}
\right.
\end{eqnarray}
\end{lemma}
\textbf{Proof:} With Lemma \ref{first} and Eq.(\ref{5.2.1}), we can easily get the results.\hfill$\Box$

\begin{lemma}\label{second}
For $i=0,1$,
\begin{enumerate}
\item if $x=x_a$ is the solution of the equation $x^{p^2}+x=-a^p$ in $\mathbb{F}_q$, $\delta_{1,i}+\delta_{3,i}$ is given by the following four cases.
    \begin{enumerate}
    \item [(1)] if $\mathrm{Tr}(x_a^{p+1})=0$, $\mathrm{Tr}(x_a)=0$,
    \begin{equation}\label{5.2.13}
    \delta_{1,i}+\delta_{3,i}=p\delta_{1,i},
    \end{equation}
    \item [(2)] if $\mathrm{Tr}(x_a^{p+1})=0$, $\mathrm{Tr}(x_a)\ne0$,
    \begin{equation}\label{5.2.14}
    \delta_{1,i}+\delta_{3,i}=0,
    \end{equation}
    \item [(3)] if $\mathrm{Tr}(x_a^{p+1})\ne0$, $\mathrm{Tr}(x_a)=0$,
    \begin{eqnarray}\nonumber
    \delta_{1,i}+\delta_{3,i}&=& \left\{
    \begin{array}{ll}
    -\eta\left(-\mathrm{Tr}(x_a^{p+1})\right)\frac{p-1}{2}G^{e+1}, & 2\nmid e, p\mid e, \\
    \eta\left(-\mathrm{Tr}(x_a^{p+1})\right)\frac{(-1)^i\eta(-e)p+1}{2}G^{e+1}, & 2\nmid e, p\nmid e, \\
    -(-1)^i\eta\left(-\mathrm{Tr}(x_a^{p+1})\right)\frac{p-1}{2}p^{\frac{e}{2}+1}, & 2\mid e,\,e\equiv 2 \bmod 4,\,p\mid e, \\
    \frac{(-1)^i+\eta(e)}{2}\eta\left(-\mathrm{Tr}(x_a^{p+1})\right)p^{\frac{e}{2}+1}, & 2\mid e,\,e\equiv 2 \bmod 4,\,p\nmid e, \\
     -(-1)^i\eta\left(-\mathrm{Tr}(x_a^{p+1})\right)\frac{p-1}{2}p^{\frac{e}{2}+2}, & 2\mid e,\,e\equiv 0 \bmod 4,\,p\mid e, \\
     \frac{(-1)^i+\eta(e)}{2}\eta\left(-\mathrm{Tr}(x_a^{p+1})\right)p^{\frac{e}{2}+2}, & 2\mid e,\,e\equiv 0 \bmod 4,\,p\nmid e, \\
    \end{array}
    \right.\\\label{5.2.15}
    \end{eqnarray}
    \item [(4)] if $\mathrm{Tr}(x_a^{p+1})\ne0$, $\mathrm{Tr}(x_a)\ne0$,
    \begin{eqnarray}\nonumber
    \delta_{1,i}+\delta_{3,i}&=& \left\{
    \begin{array}{ll}
    \frac{(-1)^i\eta(-1)p+\eta\left(-\mathrm{Tr}(x_a^{p+1})\right)}{2}G^{e+1}, \qquad \qquad 2\nmid e, p\mid e, \\
    -\eta(-e)\frac{p-1}{2}G^{e+1},\\ \qquad\qquad\qquad\qquad\qquad\qquad 2\nmid e, p\nmid e, e\mathrm{Tr}(x_a^{p+1})=\mathrm{Tr}(x_a)^2,\\
    \frac{(-1)^i\eta\left(e\mathrm{Tr}(x_a^{p+1})-\mathrm{Tr}(x_a)^2\right)p+\eta\left(-\mathrm{Tr}(x_a^{p+1})\right)}{2}G^{e+1},\\ \qquad\qquad\qquad\qquad\qquad\qquad 2\nmid e, p\nmid e, e\mathrm{Tr}(x_a^{p+1})\ne\mathrm{Tr}(x_a)^2,\\
    \frac{1+(-1)^i\eta\left(-\mathrm{Tr}(x_a^{p+1})\right)}{2}p^{\frac{e}{2}+1}, \qquad\qquad 2\mid e, e\equiv 2 \bmod 4, p\mid e, \\
    -(-1)^i\eta(-e)\frac{p-1}{2}p^{\frac{e}{2}+1},\\ \qquad\qquad\qquad\qquad\qquad 2\mid e, e\equiv 2 \bmod 4, p\nmid e, e\mathrm{Tr}(x_a^{p+1})=\mathrm{Tr}(x_a)^2,\\
    \frac{\eta\left(\mathrm{Tr}(x_a)^2-e\mathrm{Tr}(x_a^{p+1})\right)+(-1)^i\eta\left(-\mathrm{Tr}(x_a^{p+1})\right)}{2}p^{\frac{e}{2}+1},\\ \qquad\qquad\qquad\qquad\qquad 2\mid e, e\equiv 2 \bmod 4, p\nmid e, e\mathrm{Tr}(x_a^{p+1})\ne\mathrm{Tr}(x_a)^2,\\
    \frac{1+(-1)^i\eta\left(-\mathrm{Tr}(x_a^{p+1})\right)}{2}p^{\frac{e}{2}+2}, \qquad\qquad 2\mid e, e\equiv 0 \bmod 4, p\mid e, \\
    -(-1)^i\eta(-e)\frac{p-1}{2}p^{\frac{e}{2}+2},\\ \qquad\qquad\qquad\qquad\qquad 2\mid e, e\equiv 0 \bmod 4, p\nmid e, e\mathrm{Tr}(x_a^{p+1})=\mathrm{Tr}(x_a)^2,\\
    \frac{\eta\left(\mathrm{Tr}(x_a)^2-e\mathrm{Tr}(x_a^{p+1})\right)+(-1)^i\eta\left(-\mathrm{Tr}(x_a^{p+1})\right)}{2}p^{\frac{e}{2}+2},\\ \qquad\qquad\qquad\qquad\qquad 2\mid e, e\equiv 0 \bmod 4, p\nmid e, e\mathrm{Tr}(x_a^{p+1})\ne\mathrm{Tr}(x_a)^2,\\
    \end{array}
    \right.\\\label{5.2.16}
    \end{eqnarray}
    \end{enumerate}
\item if $2\mid e$ and $e\equiv 0 \bmod 4$, there are no solutions of the equation $x^{p^2}+x=-a^p$ in $\mathbb{F}_q$, then $\delta_{i3}=0$.
\end{enumerate}
\end{lemma}
\textbf{Proof:} With the definition of Weil sums, we have
\begin{eqnarray}
\delta_{3,i}&=&\sum_{c \in C_{i}^{(2,p)}}\sum_{y \in \mathbb{F}_p^*}\zeta_p^{-cy}\sum_{z \in \mathbb{F}_p^*}\sum_{x \in \mathbb{F}_q}\zeta_p^{\mathrm{Tr}(yx^{p+1}+(az-y)x)}\nonumber\\
&=&\sum_{c \in C_{i}^{(2,p)}}\sum_{y \in \mathbb{F}_p^*}\zeta_p^{-cy}\sum_{z \in \mathbb{F}_p^*}S(y,az-y).\label{5.2.17}
\end{eqnarray}
\indent if $2\nmid e$, or $2\mid e$, and $e\equiv 2\bmod 4$, $f(x)=y^px^{p^2}+yx$($y \in \mathbb{F}_p^*$) and $g(x)=x^{p^2}+x$
are all permutation polynomials over $\mathbb{F}_{q}$, and $x=x_a$ is the unique solution of the equation $x^{p^2}+x=-a^p$ in $\mathbb{F}_q$, then $x=y^{-1}zx_a+\frac{1}{2}$ is the unique solution of the equation $y^px^{p^2}+yx=-(az-y)^p$ in $\mathbb{F}_q$.\\
\indent if $2\mid e$ and $e\equiv 0 \bmod 4$,  $f(x)=y^px^{p^2}+yx$($y \in \mathbb{F}_p^*$) and $g(x)=x^{p^2}+x$ is not permutation polynomials in $\mathbb{F}_q$. There are the following two cases:
\begin{enumerate}
\item [(I)] if $x=x_a$ is the solution of the equation $x^{p^2}+x=-a^p$ in $\mathbb{F}_q$, then $x=y^{-1}zx_a+\frac{1}{2}$ is the solution of the equation  $y^px^{p^2}+yx=-(az-y)^p$ in $\mathbb{F}_q$.
\item [(II)] if there are no solutions of the equation $x^{p^2}+x=-a^p$ in $\mathbb{F}_q$, then the equation $y^px^{p^2}+yx=-(az-y)^p$ has no solutions in $\mathbb{F}_q$, thus $S(y,az-y)=0$, we can deduce $\delta_{3,i}=0$, and finish the second part of the proof.
\end{enumerate}

Next, we only need to proof part (I).\\
With Lemma \ref{weil sums5}, Lemma \ref{weil sums2} and Lemma \ref{change}, we have
\begin{eqnarray}\label{5.2.18}
S(y,az-y)&=&\left\{
\begin{array}{ll}
G^e\eta(y)\zeta_p^{\mathrm{Tr}\left(-y(y^{-1}zx_a+\frac{1}{2})^{p+1}\right)}, & 2\nmid e,\\
-p^{\frac{e}{2}}\zeta_p^{\mathrm{Tr}\left(-y(y^{-1}zx_a+\frac{1}{2})^{p+1}\right)}, & 2\mid e, e\equiv 2 \bmod 4,\\
-p^{\frac{e}{2}+1}\zeta_p^{\mathrm{Tr}\left(-y(y^{-1}zx_a+\frac{1}{2})^{p+1}\right)}, & 2\mid e, e\equiv 0 \bmod 4,\\
\end{array}
\right.
\end{eqnarray}
then
\begin{eqnarray}\label{5.2.20}
S(y,az-y)&=&\left\{
\begin{array}{ll}
G^e\eta(y)\zeta_p^{-\frac{e}{4}y}\zeta_p^{-\frac{\mathrm{Tr}(x_a^{p+1})}{y}z^2-\mathrm{Tr}(x_a)z}, & 2\nmid e,\\
-p^{\frac{e}{2}}\zeta_p^{-\frac{e}{4}y}\zeta_p^{-\frac{\mathrm{Tr}(x_a^{p+1})}{y}z^2-\mathrm{Tr}(x_a)z}, & 2\mid e, e\equiv 2 \bmod 4,\\
-p^{\frac{e}{2}+1}\zeta_p^{-\frac{e}{4}y}\zeta_p^{-\frac{\mathrm{Tr}(x_a^{p+1})}{y}z^2-\mathrm{Tr}(x_a)z}, & 2\mid e, e\equiv 0 \bmod 4,\\
\end{array}
\right.
\end{eqnarray}
Applying the values of Eq.(\ref{5.2.20}) into Eq.(\ref{5.2.17}), we have
\begin{eqnarray*}
\delta_{3,i}&=&\left\{
\begin{array}{ll}
G^e\sum\limits_{y \in\mathbb{F}_p^*}\eta(y)\zeta_p^{-\frac{e}{4}y}\sum\limits_{c\in C_i^{(2,p)}}\zeta_p^{-cy}\sum\limits_{z\in\mathbb{F}_p^*}\zeta_p^{-\frac{\mathrm{Tr}(x_a^{p+1})}{y}z^2-\mathrm{Tr}(x_a)z}, & 2\nmid e,\\
-p^{\frac{e}{2}}\sum\limits_{y \in\mathbb{F}_p^*}\zeta_p^{-\frac{e}{4}y}\sum\limits_{c\in C_i^{(2,p)}}\zeta_p^{-cy}\sum\limits_{z\in\mathbb{F}_p^*}\zeta_p^{-\frac{\mathrm{Tr}(x_a^{p+1})}{y}z^2-\mathrm{Tr}(x_a)z}, & 2\mid e, e\equiv 2 \bmod 4,\\
-p^{\frac{e}{2}+1}\sum\limits_{y \in\mathbb{F}_p^*}\zeta_p^{-\frac{e}{4}y}\sum\limits_{c\in C_i^{(2,p)}}\zeta_p^{-cy}\sum\limits_{z\in\mathbb{F}_p^*}\zeta_p^{-\frac{\mathrm{Tr}(x_a^{p+1})}{y}z^2-\mathrm{Tr}(x_a)z}, & 2\mid e, e\equiv 0 \bmod 4,\\
\end{array}
\right.
\end{eqnarray*}
With Eq.(\ref{5.2.0}),
\begin{eqnarray}\nonumber
\delta_{1,i}+\delta_{3,i}&=&\left\{
\begin{array}{ll}
G^e\sum\limits_{y \in\mathbb{F}_p^*}\eta(y)\zeta_p^{-\frac{e}{4}y}\sum\limits_{c\in C_i^{(2,p)}}\zeta_p^{-cy}\sum\limits_{z\in\mathbb{F}_p}\zeta_p^{-\frac{\mathrm{Tr}(x_a^{p+1})}{y}z^2-\mathrm{Tr}(x_a)z}, & 2\nmid e,\\
-p^{\frac{e}{2}}\sum\limits_{y \in\mathbb{F}_p^*}\zeta_p^{-\frac{e}{4}y}\sum\limits_{c\in C_i^{(2,p)}}\zeta_p^{-cy}\sum\limits_{z\in\mathbb{F}_p}\zeta_p^{-\frac{\mathrm{Tr}(x_a^{p+1})}{y}z^2-\mathrm{Tr}(x_a)z}, & 2\mid e, e\equiv 2 \bmod 4,\\
-p^{\frac{e}{2}+1}\sum\limits_{y \in\mathbb{F}_p^*}\zeta_p^{-\frac{e}{4}y}\sum\limits_{c\in C_i^{(2,p)}}\zeta_p^{-cy}\sum\limits_{z\in\mathbb{F}_p}\zeta_p^{-\frac{\mathrm{Tr}(x_a^{p+1})}{y}z^2-\mathrm{Tr}(x_a)z}, & 2\mid e, e\equiv 0 \bmod 4,\\
\end{array}
\right.\\\label{5.2.23}
\end{eqnarray}
\begin{enumerate}
\item [(1)] if $\mathrm{Tr}(x_a^{p+1})=0$, $\mathrm{Tr}(x_a)=0$, by the orthogonal property of additive character, with Eq.(\ref{5.2.0}), we have
    \begin{equation*}
    \delta_{1,i}+\delta_{3,i}=p\delta_{1,i}.
    \end{equation*}
\item [(2)] if $\mathrm{Tr}(x_a^{p+1})=0$, $\mathrm{Tr}(x_a)\ne0$, by the orthogonal property of additive character, we have
    \begin{equation*}
    \delta_{1,i}+\delta_{3,i}=0.
    \end{equation*}
\item [(3)] if $\mathrm{Tr}(x_a^{p+1})\ne0$, $\mathrm{Tr}(x_a)=0$,
with Lemma \ref{quadratic sums}, we have
\begin{eqnarray*}
\delta_{1,i}+\delta_{3,i}&=&\left\{
\begin{array}{ll}
\eta\left(-\mathrm{Tr}(x_a^{p+1})\right)G^{e+1}\sum\limits_{y \in\mathbb{F}_p^*}\zeta_p^{-\frac{e}{4}y}\sum\limits_{c\in C_i^{(2,p)}}\zeta_p^{-cy}, & 2\nmid e,\\
-\eta\left(\mathrm{Tr}(x_a^{p+1})\right)Gp^{\frac{e}{2}}\sum\limits_{y \in\mathbb{F}_p^*}\zeta_p^{-\frac{e}{4}y}\eta(-y)\sum\limits_{c\in C_i^{(2,p)}}\zeta_p^{-cy}, & 2\mid e, e\equiv 2 \bmod 4,\\
-\eta\left(\mathrm{Tr}(x_a^{p+1})\right)Gp^{\frac{e}{2}+1}\sum\limits_{y \in\mathbb{F}_p^*}\zeta_p^{-\frac{e}{4}y}\eta(-y)\sum\limits_{c\in C_i^{(2,p)}}\zeta_p^{-cy}, & 2\mid e, e\equiv 0 \bmod 4,\\
\end{array}
\right.
\end{eqnarray*}
if $p\mid e$, $\zeta_p^{-\frac{ey}{4}}=1$. With the definition of the quadratic Gaussian period, we have
\begin{eqnarray*}
\delta_{1,i}+\delta_{3,i}&=&\left\{
\begin{array}{ll}
\eta\left(-\mathrm{Tr}(x_a^{p+1})\right)G^{e+1}\sum\limits_{c\in C_i^{(2,p)}}\big(\sum\limits_{y \in\mathbb{F}_p^*}\zeta_p^{-cy}\big),\qquad\qquad\qquad\qquad\qquad 2\nmid e, p\mid e,\\
\eta\left(-\mathrm{Tr}(x_a^{p+1})\right)G^{e+1}\big(\sum\limits_{-y \in C_i^{(2,p)}}\zeta_p^{-\frac{e}{4}y}\rho_0+\sum\limits_{-y \in C_{1-i}^{(2,p)}}\zeta_p^{-\frac{e}{4}y}\rho_1\big),~\qquad 2\nmid e, p\nmid e,\\
-\eta\left(\mathrm{Tr}(x_a^{p+1})\right)Gp^{\frac{e}{2}}\big(\sum\limits_{-y \in C_i^{(2,p)}}\eta(-y)\rho_0+\sum\limits_{-y \in C_{1-i}^{(2,p)}}\eta(-y)\rho_1\big),\\ \qquad\qquad\qquad\qquad\qquad\qquad\qquad\qquad\qquad\qquad\qquad 2\mid e, e\equiv 2 \bmod 4, p\mid e,\\
-\eta\left(\mathrm{Tr}(x_a^{p+1})\right)Gp^{\frac{e}{2}}\big(\sum\limits_{-y \in C_i^{(2,p)}}\zeta_p^{-\frac{e}{4}y}\eta(-y)\rho_0+\sum\limits_{-y \in C_{1-i}^{(2,p)}}\zeta_p^{-\frac{e}{4}y}\eta(-y)\rho_1\big),\\ \qquad\qquad\qquad\qquad\qquad\qquad\qquad\qquad\qquad\qquad\qquad 2\mid e, e\equiv 2 \bmod 4, p\nmid e,\\
-\eta\left(\mathrm{Tr}(x_a^{p+1})\right)Gp^{\frac{e}{2}+1}\big(\sum\limits_{-y \in C_i^{(2,p)}}\eta(-y)\rho_0+\sum\limits_{-y \in C_{1-i}^{(2,p)}}\eta(-y)\rho_1\big),\\ \qquad\qquad\qquad\qquad\qquad\qquad\qquad\qquad\qquad\qquad\qquad 2\mid e, e\equiv 0 \bmod 4, p\mid e,\\
-\eta\left(\mathrm{Tr}(x_a^{p+1})\right)Gp^{\frac{e}{2}+1}\big(\sum\limits_{-y \in C_i^{(2,p)}}\zeta_p^{-\frac{e}{4}y}\eta(-y)\rho_0+\sum\limits_{-y \in C_{1-i}^{(2,p)}}\zeta_p^{-\frac{e}{4}y}\eta(-y)\rho_1\big),\\ \qquad\qquad\qquad\qquad\qquad\qquad\qquad\qquad\qquad\qquad\qquad 2\mid e, e\equiv 0 \bmod 4, p\nmid e,\\
\end{array}
\right.
\end{eqnarray*}
by the orthogonal property of additive character,
\begin{eqnarray*}
\delta_{1,i}+\delta_{3,i}&=&\left\{
\begin{array}{ll}
-\eta\left(-\mathrm{Tr}(x_a^{p+1})\right)\frac{p-1}{2}G^{e+1}, & 2\nmid e, p\mid e,\\
\eta\left(-\mathrm{Tr}(x_a^{p+1})\right)G^{e+1}(\rho_0^2+\rho_1^2),& 2\nmid e, e\in C_i^{(2,p)},\\
\eta\left(-\mathrm{Tr}(x_a^{p+1})\right)G^{e+1}(2\rho_0\rho_1), & 2\nmid e, e\in C_{1-i}^{(2,p)},\\
-(-1)^i\eta\left(\mathrm{Tr}(x_a^{p+1})\right)\frac{p-1}{2}Gp^{\frac{e}{2}}(\rho_0-\rho_1), & 2\mid e, e\equiv 2 \bmod 4, p\mid e,\\
-(-1)^i\eta\left(\mathrm{Tr}(x_a^{p+1})\right)Gp^{\frac{e}{2}}(\rho_0^2-\rho_1^2), & 2\mid e, e\equiv 2 \bmod 4, e\in C_i^{(2,p)},\\
-(-1)^i\eta\left(\mathrm{Tr}(x_a^{p+1})\right)Gp^{\frac{e}{2}}(\rho_1\rho_0-\rho_0\rho_1), & 2\mid e, e\equiv 2 \bmod 4, e\in C_{1-i}^{(2,p)},\\
-(-1)^i\eta\left(\mathrm{Tr}(x_a^{p+1})\right)\frac{p-1}{2}Gp^{\frac{e}{2}+1}(\rho_0-\rho_1), & 2\mid e, e\equiv 0 \bmod 4, p\mid e,\\
-(-1)^i\eta\left(\mathrm{Tr}(x_a^{p+1})\right)Gp^{\frac{e}{2}+1}(\rho_0^2-\rho_1^2), & 2\mid e, e\equiv 0 \bmod 4, e\in C_i^{(2,p)},\\
-(-1)^i\eta\left(\mathrm{Tr}(x_a^{p+1})\right)Gp^{\frac{e}{2}+1}(\rho_1\rho_0-\rho_0\rho_1), & 2\mid e, e\equiv 0 \bmod 4, e\in C_{1-i}^{(2,p)},\\
\end{array}
\right.
\end{eqnarray*}
with Lemma \ref{Gaussian period}, we have
\begin{eqnarray*}
\delta_{1,i}+\delta_{3,i}&=&\left\{
\begin{array}{ll}
-\eta\left(-\mathrm{Tr}(x_a^{p+1})\right)\frac{p-1}{2}G^{e+1}, & 2\nmid e, p\mid e,\\
\eta\left(-\mathrm{Tr}(x_a^{p+1})\right)\frac{1+\eta(-1)p}{2}G^{e+1},& 2\nmid e, e\in C_i^{(2,p)},\\
\eta\left(-\mathrm{Tr}(x_a^{p+1})\right)\frac{1-\eta(-1)p}{2}G^{e+1}, & 2\nmid e, e\in C_{1-i}^{(2,p)},\\
-(-1)^i\eta\left(-\mathrm{Tr}(x_a^{p+1})\right)\frac{p-1}{2}p^{\frac{e}{2}+1}, & 2\mid e, e\equiv 2 \bmod 4, p\mid e,\\
(-1)^i\eta\left(-\mathrm{Tr}(x_a^{p+1})\right)p^{\frac{e}{2}+1}, & 2\mid e, e\equiv 2 \bmod 4, e\in C_i^{(2,p)},\\
0, & 2\mid e, e\equiv 2 \bmod 4, e\in C_{1-i}^{(2,p)},\\
-(-1)^i\eta\left(-\mathrm{Tr}(x_a^{p+1})\right)\frac{p-1}{2}p^{\frac{e}{2}+2}, & 2\mid e, e\equiv 0 \bmod 4, p\mid e,\\
(-1)^i\eta\left(-\mathrm{Tr}(x_a^{p+1})\right)p^{\frac{e}{2}+2}, & 2\mid e, e\equiv 0 \bmod 4, e\in C_i^{(2,p)},\\
0, & 2\mid e, e\equiv 0 \bmod 4, e\in C_{1-i}^{(2,p)},\\
\end{array}
\right.
\end{eqnarray*}
After simplification, we can get Eq.(\ref{5.2.15}).
\item [(4)] if $\mathrm{Tr}(x_a^{p+1})\ne0$, $\mathrm{Tr}(x_a)\ne0$,
With Lemma \ref{quadratic sums}, we have
\begin{eqnarray*}
\delta_{1,i}+\delta_{3,i}&=&\left\{
\begin{array}{ll}
\eta\left(-\mathrm{Tr}(x_a^{p+1})\right)G^{e+1}\sum\limits_{y \in\mathbb{F}_p^*}\zeta_p^{\frac{e\mathrm{Tr}(x_a^{p+1})-\mathrm{Tr}(x_a)^2}{4\mathrm{Tr}(x_a^{p+1})}\cdot(-y)}\sum\limits_{c\in C_i^{(2,p)}}\zeta_p^{-cy}, \qquad 2\nmid e,\\
-\eta\left(\mathrm{Tr}(x_a^{p+1})\right)Gp^{\frac{e}{2}}\sum\limits_{y \in\mathbb{F}_p^*}\zeta_p^{\frac{e\mathrm{Tr}(x_a^{p+1})-\mathrm{Tr}(x_a)^2}{4\mathrm{Tr}(x_a^{p+1})}\cdot(-y)}\eta(-y)\sum\limits_{c\in C_i^{(2,p)}}\zeta_p^{-cy},\\ \qquad\qquad\qquad\qquad\qquad\qquad\qquad\qquad\qquad\qquad 2\mid e, e\equiv 2 \bmod 4,\\
-\eta\left(\mathrm{Tr}(x_a^{p+1})\right)Gp^{\frac{e}{2}+1}\sum\limits_{y \in\mathbb{F}_p^*}\zeta_p^{\frac{e\mathrm{Tr}(x_a^{p+1})-\mathrm{Tr}(x_a)^2}{4\mathrm{Tr}(x_a^{p+1})}\cdot(-y)}\eta(-y)\sum\limits_{c\in C_i^{(2,p)}}\zeta_p^{-cy},\\ \qquad\qquad\qquad\qquad\qquad\qquad\qquad\qquad\qquad\qquad 2\mid e, e\equiv 0 \bmod 4,\\
\end{array}
\right.
\end{eqnarray*}
With the similar discussion of Lemma \ref{first} and part (3) in this proof, we can get
\begin{eqnarray*}\nonumber
\delta_{1,i}+\delta_{3,i}&=&\left\{
\begin{array}{ll}
(-1)^i\frac{1+\eta(-1)p}{2}G^{e+1}, ~\qquad\qquad\qquad 2\nmid e, p\mid e,\left(-\mathrm{Tr}(x_a^{p+1})\right) \in C_i^{(2,p)},\\
-(-1)^i\frac{1-\eta(-1)p}{2}G^{e+1}, ~\quad\qquad\qquad 2\nmid e, p\mid e,\left(-\mathrm{Tr}(x_a^{p+1})\right) \in C_{1-i}^{(2,p)},\\
-\eta\left(-\mathrm{Tr}(x_a^{p+1})\right)\frac{p-1}{2}G^{e+1}, ~\quad\qquad 2\nmid e, p\nmid e, e\mathrm{Tr}(x_a^{p+1})=\mathrm{Tr}(x_a)^2,\\
\eta\left(-\mathrm{Tr}(x_a^{p+1})\right)\frac{1+\eta(-1)p}{2}G^{e+1}, ~\qquad 2\nmid e, p\nmid e, \frac{e\mathrm{Tr}(x_a^{p+1})-\mathrm{Tr}(x_a)^2}{\mathrm{Tr}(x_a^{p+1})} \in C_i^{(2,p)},\\
\eta\left(-\mathrm{Tr}(x_a^{p+1})\right)\frac{1-\eta(-1)p}{2}G^{e+1}, ~\qquad 2\nmid e, p\nmid e, \frac{e\mathrm{Tr}(x_a^{p+1})-\mathrm{Tr}(x_a)^2}{\mathrm{Tr}(x_a^{p+1})} \in C_{1-i}^{(2,p)},\\
p^{\frac{e}{2}+1}, \qquad\qquad\qquad\qquad 2\mid e, e\equiv2\bmod4, p\mid e, \left(-\mathrm{Tr}(x_a^{p+1})\right) \in C_i^{(2,p)},\\
0, ~\qquad\qquad\qquad\qquad\quad 2\mid e, e\equiv2\bmod4, p\mid e, \left(-\mathrm{Tr}(x_a^{p+1})\right) \in C_{1-i}^{(2,p)},\\
-(-1)^i\eta\left(-\mathrm{Tr}(x_a^{p+1})\right)\frac{p-1}{2}p^{\frac{e}{2}+1},\\ \qquad\qquad\qquad\qquad\qquad 2\mid e, e\equiv2\bmod4, p\nmid e, e\mathrm{Tr}(x_a^{p+1})=\mathrm{Tr}(x_a)^2,\\
(-1)^i\eta\left(-\mathrm{Tr}(x_a^{p+1})\right)p^{\frac{e}{2}+1},\\ \qquad\qquad\qquad\qquad\qquad 2\mid e, e\equiv2\bmod4, p\nmid e, \frac{e\mathrm{Tr}(x_a^{p+1})-\mathrm{Tr}(x_a)^2}{\mathrm{Tr}(x_a^{p+1})} \in C_i^{(2,p)},\\
0, \qquad\qquad\qquad\qquad\quad 2\mid e, e\equiv2\bmod4, p\nmid e, \frac{e\mathrm{Tr}(x_a^{p+1})-\mathrm{Tr}(x_a)^2}{\mathrm{Tr}(x_a^{p+1})} \in C_{1-i}^{(2,p)},\\
\end{array}
\right.
\end{eqnarray*}
\begin{eqnarray}\nonumber
\delta_{1,i}+\delta_{3,i}&=&\left\{
\begin{array}{ll}
p^{\frac{e}{2}+2}, ~~\quad\qquad\qquad\qquad 2\mid e, e\equiv0\bmod4, p\mid e, \left(-\mathrm{Tr}(x_a^{p+1})\right) \in C_i^{(2,p)},\\
0, ~\qquad\qquad\qquad\qquad\quad 2\mid e, e\equiv0\bmod4, p\mid e, \left(-\mathrm{Tr}(x_a^{p+1})\right) \in C_{1-i}^{(2,p)},\\
-(-1)^i\eta\left(-\mathrm{Tr}(x_a^{p+1})\right)\frac{p-1}{2}p^{\frac{e}{2}+2},\\ \qquad\qquad\qquad\qquad\qquad 2\mid e, e\equiv0\bmod4, p\nmid e, e\mathrm{Tr}(x_a^{p+1})=\mathrm{Tr}(x_a)^2,\\
(-1)^i\eta\left(-\mathrm{Tr}(x_a^{p+1})\right)p^{\frac{e}{2}+2},\\ \qquad\qquad\qquad\qquad\qquad 2\mid e, e\equiv0\bmod4, p\nmid e, \frac{e\mathrm{Tr}(x_a^{p+1})-\mathrm{Tr}(x_a)^2}{\mathrm{Tr}(x_a^{p+1})} \in C_i^{(2,p)},\\
0, \qquad\qquad\qquad\qquad\quad 2\mid e, e\equiv0\bmod4, p\nmid e, \frac{e\mathrm{Tr}(x_a^{p+1})-\mathrm{Tr}(x_a)^2}{\mathrm{Tr}(x_a^{p+1})} \in C_{1-i}^{(2,p)},\\
\end{array}
\right.\\\label{5.2.31}
\end{eqnarray}
After simplification, we can deduce Eq.(\ref{5.2.16}).\hfill$\Box$
\end{enumerate}

Applying Lemma \ref{first} and Lemma \ref{second} into Eq.(\ref{5.2.7}), we can easily get the Hamming weight of the codes $\mathcal{C}_{D_{2i}}$($i=0,1$). In order to calculate the frequency $A_{wt_{ij}}$($i=0,1$) of the linear codes, we need to show the following lemmas at first.

\begin{lemma}\label{F1}(\cite{CCK})
Let
\begin{eqnarray*}
N_0&=&\left|\{x \in \mathbb{F}_q : \mathrm{Tr}(x^{p+1})=0\}\right|,\nonumber\\
N_{(0,0)}&=&\left|\{x \in \mathbb{F}_q : \mathrm{Tr}(x^{p+1})=0, \mathrm{Tr}(x)=0\}\right|,\label{5.2.32}\\
N_{(0,\overline{0})}&=&\left|\{x \in \mathbb{F}_q : \mathrm{Tr}(x^{p+1})=0, \mathrm{Tr}(x)\ne0\}\right|,\nonumber
\end{eqnarray*}
then
\begin{enumerate}
\item
\begin{eqnarray*}\label{5.2.33}
N_{(0,0)}&=&\left\{
\begin{array}{ll}
p^{e-2}, & 2\nmid e, p \mid e,\\
p^{e-2}+\eta(-e)\frac{p-1}{p^2}G^{e+1}, & 2\nmid e, p\nmid e,\\
p^{e-2}-p^{\frac{e}{2}-1}(p-1), & 2\mid e, e\equiv 2 \bmod4, p\mid e,\\
p^{e-2}-p^{\frac{e}{2}}(p-1), & 2\mid e, e\equiv 0 \bmod4, p\mid e,\\
p^{e-2}, & 2\mid e, p\nmid e.
\end{array}
\right.
\end{eqnarray*}
\item
\begin{eqnarray*}\label{5.2.34}
N_{(0,\overline{0})}&=&\left\{
\begin{array}{ll}
p^{e-2}(p-1), & p\mid e,\\
p^{e-2}(p-1)-\eta(-e)\frac{p-1}{p^2}G^{e+1}, & 2\nmid e, p\nmid e,\\
p^{e-2}(p-1)-p^{\frac{e}{2}-1}(p-1), & 2\mid e, e\equiv 2 \bmod 4, p\nmid e,\\
p^{e-2}(p-1)-p^{\frac{e}{2}}(p-1), & 2\mid e, e\equiv 0 \bmod 4, p\nmid e.\\
\end{array}
\right.
\end{eqnarray*}
\end{enumerate}
\end{lemma}

\begin{lemma}\label{F2}
For $l \in \{0,1\}$, let
\begin{eqnarray*}
M_{(l,0)}&=&\left|\{x \in \mathbb{F}_q : \left(\frac{\mathrm{Tr}(x^{p+1})}{p}\right)=(-1)^l, \mathrm{Tr}(x)=0\}\right|,\label{5.2.35}\\
M_{(l,\overline{0})}&=&\left|\{x \in \mathbb{F}_q : \left(\frac{\mathrm{Tr}(x^{p+1})}{p}\right)=(-1)^l, \mathrm{Tr}(x)\ne0\}\right|,\label{5.2.36}
\end{eqnarray*}
then
\begin{enumerate}
\item
\begin{eqnarray*}\label{5.2.37}
M_{(l,0)}&=&\left\{
\begin{array}{ll}
\frac{p-1}{2}p^{e-2}+(-1)^l\eta(-1)\frac{p-1}{2p}G^{e+1}, & 2\nmid e, p\mid e,\\
\frac{p-1}{2}p^{e-2}-\eta(-e)\frac{p-1}{2p^2}G^{e+1}, & 2\nmid e, p\nmid e,\\
\frac{p-1}{2}p^{e-2}+\frac{p-1}{2}p^{\frac{e}{2}-1}, & 2\mid e, e\equiv2\bmod4, p\mid e,\\
\frac{p-1}{2}p^{e-2}-(-1)^l\eta(-e)\frac{p-1}{2}p^{\frac{e}{2}-1}, & 2\mid e, e\equiv2\bmod4, p\nmid e,\\
\frac{p-1}{2}p^{e-2}+\frac{p-1}{2}p^{\frac{e}{2}}, & 2\mid e, e\equiv0\bmod4, p\mid e,\\
\frac{p-1}{2}p^{e-2}-(-1)^l\eta(-e)\frac{p-1}{2}p^{\frac{e}{2}}, & 2\mid e, e\equiv0\bmod4, p\nmid e,\\
\end{array}
\right.
\end{eqnarray*}
\item
\begin{eqnarray*}\label{5.2.38}
M_{(l,\overline{0})}&=&\left\{
\begin{array}{ll}
\frac{(p-1)^2}{2}p^{e-2}, & p\mid e,\\
\frac{(p-1)^2}{2}p^{e-2}+\left((-1)^l\eta(-1)p+\eta(-e)\right)\frac{p-1}{2p^2}G^{e+1}, & 2\nmid e, p\nmid e,\\
\frac{(p-1)^2}{2}p^{e-2}+\left(1+(-1)^l\eta(-e)\right)\frac{p-1}{2}p^{\frac{e}{2}-1}, & 2\mid e, e \equiv2\bmod4, p\nmid e,\\
\frac{(p-1)^2}{2}p^{e-2}+\left(1+(-1)^l\eta(-e)\right)\frac{p-1}{2}p^{\frac{e}{2}}, & 2\mid e, e \equiv0\bmod4, p\nmid e,\\
\end{array}
\right.
\end{eqnarray*}
\end{enumerate}
\end{lemma}
\textbf{Proof:} By the orthogonal property of additive character, we have
\begin{eqnarray}
M_{(l,0)}&=&\sum\limits_{c \in C_l^{(2,p)}}\sum\limits_{x \in \mathbb{F}_q}\left(\frac{1}{p}\sum\limits_{y \in \mathbb{F}_p}\zeta_p^{y\left(\mathrm{Tr}(x^{p+1})-c\right)}\right)\left(\frac{1}{p}\sum\limits_{z \in \mathbb{F}_p}\zeta_p^{z\mathrm{Tr}(x)}\right)\nonumber\\
&=&p^{-2}\sum\limits_{c \in C_l^{(2,p)}}\sum\limits_{x \in \mathbb{F}_q}\left(1+\sum\limits_{y \in \mathbb{F}_p^*}\zeta_p^{y\left(\mathrm{Tr}(x^{p+1})-c\right)}\right)\left(1+\sum\limits_{z \in \mathbb{F}_p^*}\zeta_p^{z\mathrm{Tr}(x)}\right)\nonumber\\
&=&\frac{p-1}{2}p^{e-2}+p^{-2}(\xi_{l1}+\xi_{l2}+\xi_{l3}),\label{5.2.39}
\end{eqnarray}
where,
\begin{eqnarray}
\xi_{l1}&=&\sum\limits_{c \in C_l^{(2,p)}}\sum\limits_{x \in \mathbb{F}_q}\sum\limits_{y \in \mathbb{F}_q^*}\zeta_p^{\mathrm{Tr}(yx^{p+1})-cy}\nonumber\\
&=&\sum\limits_{y \in \mathbb{F}_p^*}\sum\limits_{x \in \mathbb{F}_q}\zeta_p^{\mathrm{Tr}(yx^{p+1})}\sum\limits_{c \in C_l^{(2,p)}}\zeta_p^{-cy}\nonumber\\
&=&\sum\limits_{y \in \mathbb{F}_p^*}S(y,0)\sum\limits_{c \in C_l^{(2,p)}}\zeta_p^{-cy},\label{5.2.40}
\end{eqnarray}
With Lemma \ref{weil sums5}, Lemma \ref{weil sums1} and Lemma \ref{change},
\begin{eqnarray}\label{5.2.41}
S(y,0)=\sum\limits_{x \in \mathbb{F}_q}\zeta_p^{\mathrm{Tr}(yx^{p+1})}&=&\left\{
\begin{array}{ll}
\eta(y)G^e, & 2\nmid e,\\
-p^{\frac{e}{2}}, & e\equiv 2\bmod4,\\
-p^{\frac{e}{2}+1}, & e\equiv 0\bmod4.
\end{array}
\right.
\end{eqnarray}
Applying Eq.(\ref{5.2.41}) into Eq.(\ref{5.2.40}), we have
\begin{eqnarray}
\xi_{l1}&=&\left\{
\begin{array}{ll}\label{5.2.43}
(-1)^l\eta(-1)\frac{p-1}{2}G^{e+1}, & 2\nmid e,\\
\frac{p-1}{2}p^{\frac{e}{2}}, & 2\mid e, e\equiv2\bmod4,\\
\frac{p-1}{2}p^{\frac{e}{2}+1}, & 2\mid e, e\equiv0\bmod4.\\
\end{array}
\right.
\end{eqnarray}
By the orthogonal property of additive character, we have
\begin{equation}\label{5.2.44}
\xi_{l2}=\sum\limits_{c \in C_l^{(2,p)}}\sum\limits_{z \in \mathbb{F}_p^*}\sum\limits_{x \in \mathbb{F}_q}\zeta_p^{\mathrm{Tr}(zx)}=0.
\end{equation}
\begin{eqnarray*}
\xi_{l3}&=&\sum\limits_{y \in \mathbb{F}_p^*}\sum\limits_{z \in \mathbb{F}_p^*}\sum\limits_{x \in \mathbb{F}_q}\zeta_p^{\mathrm{Tr}(yx^{p+1}+zx)}\sum\limits_{c \in C_l^{(2,p)}}\zeta_p^{-cy},\nonumber\\
&=&\sum\limits_{y \in \mathbb{F}_p^*}\sum\limits_{z \in \mathbb{F}_p^*}S(y,z)\sum\limits_{c \in C_l^{(2,p)}}\zeta_p^{-cy},\label{5.2.45}\\\nonumber
\end{eqnarray*}
With Lemma \ref{quadratic sums} and Lemma \ref{Gaussian period}, we have
\begin{enumerate}
\item [(1)] if $p\mid e$,
\begin{equation}\label{5.2.48}
\xi_{l1}+\xi_{l3}=p\xi_{l1},
\end{equation}
\item [(2)] if $p\nmid e$,
\begin{eqnarray}\label{5.2.49}
\xi_{l1}+\xi_{l3}&=&\left\{
\begin{array}{ll}
-\eta(-e)\frac{p-1}{2}G^{e+1}, & 2\nmid e,\\
-(-1)^l\eta(-e)\frac{p-1}{2}p^{\frac{e}{2}+1}, & 2\mid e, e\equiv2\bmod4,\\
-(-1)^l\eta(-e)\frac{p-1}{2}p^{\frac{e}{2}+2}, & 2\mid e, e\equiv0\bmod4,\\
\end{array}
\right.
\end{eqnarray}
\end{enumerate}
we can get the first part of this lemma.

Let $M_{(l)}=\left|\{x \in \mathbb{F}_q : \left(\frac{\mathrm{Tr}(x^{p+1})}{p}\right)=(-1)^l \}\right|$, then
\begin{eqnarray*}
M_{(l)}&=&\sum\limits_{c \in C_l^{(2,p)}}\sum\limits_{x \in \mathbb{F}_q}\frac{1}{p}\sum\limits_{y \in \mathbb{F}_p}\zeta_p^{y(\mathrm{Tr}(x^{p+1})-c)},\\
&=&\frac{p-1}{2}p^{e-1}+\frac{1}{p}\sum\limits_{c \in C_l^{(2,p)}}\sum\limits_{y \in \mathbb{F}_p^*}\zeta_p^{-cy}\sum\limits_{x \in \mathbb{F}_q}\zeta_p^{\mathrm{Tr}(yx^{p+1})},\\
&=&\frac{p-1}{2}p^{e-1}+\frac{1}{p}\sum\limits_{c \in C_l^{(2,p)}}\sum\limits_{y \in \mathbb{F}_p^*}\zeta_p^{-cy}S(y,0),
\end{eqnarray*}
then
\begin{eqnarray*}\label{5.2.50}
M_{(l)}&=&\left\{
\begin{array}{ll}
\frac{p-1}{2}p^{e-1}+(-1)^l\eta(-1)\frac{p-1}{2p}G^{e+1}, & 2\nmid e,\\
\frac{p-1}{2}p^{e-1}+\frac{p-1}{2}p^{\frac{e}{2}-1}, & 2\mid e, e\equiv2\bmod4,\\
\frac{p-1}{2}p^{e-1}+\frac{p-1}{2}p^{\frac{e}{2}}, & 2\mid e, e\equiv0\bmod4.\\
\end{array}
\right.
\end{eqnarray*}
With $M_{(l,\overline{0})}=M_{(l)}-M_{(l,0)}$, we can easily deduce the second part of the proof.\hfill$\Box$

\begin{lemma}\label{F3}(\cite{CCK})
Let $p\nmid e$, we have
\begin{eqnarray*}
N(\overline{0},\overline{0},e)&=&\left|\{x \in \mathbb{F}_q : \mathrm{Tr}(x^{p+1})\ne 0, \mathrm{Tr}(x)\ne 0, \mathrm{Tr}(x)^2=e\mathrm{Tr}(x^{p+1}),\}\right|\nonumber\\
&=&\left\{
\begin{array}{ll}
(p-1)p^{e-2}, & 2\mid e,\\
(p-1)p^{e-2}+\eta(-e)\frac{(p-1)^2}{p^2}G^{e+1}, & 2\nmid e.\\
\end{array}
\right.\label{5.2.51}
\end{eqnarray*}
\end{lemma}

\begin{lemma}\label{F4}
Let $p\nmid e$, $s\in \mathbb{F}_p^*$ and $s\ne e$,
\begin{equation*}
V_s=\{x \in \mathbb{F}_q : \mathrm{Tr}(x^{p+1}) \ne 0, \mathrm{Tr}(x) \ne 0, \mathrm{Tr}(x)^2=s\mathrm{Tr}(x^{p+1})\},
\end{equation*}
then
\begin{eqnarray*}\label{5.2.52}
|V_s|&=&\left\{
\begin{array}{ll}
(p-1)p^{e-2}-\eta(-e)\frac{p-1}{p^2}G^{e+1}, & 2\nmid e,\\
(p-1)p^{e-2}-\eta(s)\eta(s-e)p^{\frac{e}{2}-1}(p-1), & 2\mid e, e\equiv2\bmod4,\\
(p-1)p^{e-2}-\eta(s)\eta(s-e)p^{\frac{e}{2}}(p-1), & 2\mid e, e\equiv0\bmod4.\\
\end{array}
\right.
\end{eqnarray*}
\end{lemma}
\textbf{Proof:} The proof is similar to that of Lemma \ref{F2}, so we omit the process.\hfill$\Box$

\begin{lemma}\label{F5}
Let $p\nmid e$, for $k,j \in \{0,1\}$,
\begin{equation*}
\nu(k,j)=\{s\in \mathbb{F}_p^* : s\ne e, \left(\frac{s}{p}\right)=(-1)^k, \left(\frac{s-e}{p}\right)=(-1)^j(-1)^{\frac{p-1}{2}}\},
\end{equation*}
then
\begin{equation}\label{5.2.63}
|\nu(k,j)|=\frac{1}{4}(p-2-(-1)^k\eta(e)-(-1)^j\eta(e)-(-1)^{k+j}\eta(-1)).
\end{equation}
\end{lemma}
\textbf{Proof:} With Lemma \ref{number}, we can easily deduce the result.\hfill$\Box$

\begin{lemma}\label{F6}
Let $p\nmid e$, for $k,j \in \{0,1\}$,
\begin{eqnarray}
& &N(k, \overline{0}, \overline{0}, j)\nonumber\\
&=&\left|\{x \in \mathbb{F}_q : \mathrm{Tr}(x^{p+1}) \in C_k^{(2,p)}, \mathrm{Tr}(x)\ne0, \mathrm{Tr}(x)^2\ne e\mathrm{Tr}(x^{p+1}), \frac{e\mathrm{Tr}(x^{p+1})-\mathrm{Tr}(x)^2}{\mathrm{Tr}(x^{p+1})} \in C_j^{(2,p)}\}\right|\nonumber\\
&=&\left\{
\begin{array}{ll}
\frac{1}{4}(p-2-(-1)^k\eta(e)-(-1)^j\eta(e)-(-1)^{k+j}\eta(-1))((p-1)p^{e-2}-\eta(-e)\frac{p-1}{p^2}G^{e+1}),\\ \qquad\qquad\qquad\qquad\qquad\qquad\qquad\qquad\qquad\qquad\qquad\qquad\qquad\qquad 2\nmid e,\\
\frac{1}{4}(p-2-(-1)^k\eta(e)-(-1)^j\eta(e)-(-1)^{k+j}\eta(-1))((p-1)p^{e-2}-(-1)^{k+j}\eta(-1)(p-1)p^{\frac{e}{2}-1}),\\  \qquad\qquad\qquad\qquad\qquad\qquad\qquad\qquad\qquad\qquad\qquad\qquad\qquad\qquad 2\mid e, e\equiv2 \bmod4,\\
\frac{1}{4}(p-2-(-1)^k\eta(e)-(-1)^j\eta(e)-(-1)^{k+j}\eta(-1))((p-1)p^{e-2}-(-1)^{k+j}\eta(-1)(p-1)p^{\frac{e}{2}}),\\  \qquad\qquad\qquad\qquad\qquad\qquad\qquad\qquad\qquad\qquad\qquad\qquad\qquad\qquad 2\mid e, e\equiv0 \bmod4.\\
\end{array}
\right.\nonumber\\\label{5.2.64}
\end{eqnarray}
\end{lemma}
\textbf{Proof:}
\begin{eqnarray}
& &N(k, \overline{0}, \overline{0}, j)\nonumber\\
&=&\left|\{x \in \mathbb{F}_q : \mathrm{Tr}(x)=c, \mathrm{Tr}(x^{p+1})=\frac{c^2}{s}, \forall c \in \mathbb{F}_p^*, s \in \mathbb{F}_p^*, \frac{c^2}{s} \in C_k^{(2,p)}, s\ne e, (e-s)\in C_j^{(2,p)} \}\right|\nonumber\\
&=&\left|\{x \in \mathbb{F}_q : \mathrm{Tr}(x)=c, \mathrm{Tr}(x^{p+1})=\frac{c^2}{s}, \forall c \in \mathbb{F}_p^*, s \in C_k^{(2,p)}, (e-s)\in C_j^{(2,p)} \}\right|\nonumber\\
&=&\left|\{x \in \mathbb{F}_q : \mathrm{Tr}(x)\ne 0, \mathrm{Tr}(x)^2=s\mathrm{Tr}(x^{p+1}), \forall s \in C_k^{(2,p)}, (e-s)\in C_j^{(2,p)} \}\right|\nonumber\\
&=&\left|\{x \in \mathbb{F}_q : x \in V_s, s \in \nu(k,j)\}\right|\nonumber\\
&=&\sum\limits_{s \in \nu(k,j)}|V_s|,\nonumber
\end{eqnarray}
With Lemma \ref{F4} and Lemma \ref{F5}, we can get the results.\hfill$\Box$

\textbf{Proof of Theorem \ref{weight1}:} Let $2\nmid e$ and $p\mid e$, with Lemma \ref{length}, for $i=0,1$, we have
\begin{equation*}
n_i=\frac{p-1}{2}p^{e-1}+(-1)^i\frac{p-1}{2}G^{e-1}.
\end{equation*}
Applying Lemma \ref{first} and Lemma \ref{second} into Eq.(\ref{5.2.7}), we can get the following lemma directly.
\begin{lemma}\label{w1}
Let $2\nmid e$ and $p\mid e$, for $i=0,1$, we have
\begin{eqnarray*}
wt_{i}(\mathbf{c}(a))&=&\left\{
\begin{array}{ll}
\frac{(p-1)^2}{2}p^{e-2}, ~\qquad\qquad\qquad\qquad\qquad\qquad\qquad\quad \mathrm{Tr}(x_a^{p+1})=0, \mathrm{Tr}(x_a)=0,\nonumber\\
\frac{(p-1)^2}{2}p^{e-2}+(-1)^i\frac{p-1}{2}G^{e-1}, ~\qquad\qquad\qquad\qquad \mathrm{Tr}(x_a^{p+1})=0, \mathrm{Tr}(x_a)\ne0,\nonumber\\
\frac{(p-1)^2}{2}p^{e-2}+(-1)^i\frac{p-1}{2}G^{e-1}+\eta(-\mathrm{Tr}(x_a^{p+1}))\frac{p-1}{2}G^{e-3},\nonumber\\ \qquad\qquad\qquad\qquad\qquad\qquad\qquad\qquad\qquad\qquad \mathrm{Tr}(x_a^{p+1})\ne0, \mathrm{Tr}(x_a)=0,\nonumber\\
\frac{(p-1)^2}{2}p^{e-2}+(-1)^i\frac{p-2}{2}G^{e-1}
-\frac{1}{2}\eta(-\mathrm{Tr}(x_a^{p+1}))G^{e-3},\nonumber\\ \qquad\qquad\qquad\qquad\qquad\qquad\qquad\qquad\qquad\qquad \mathrm{Tr}(x_a^{p+1})\ne0, \mathrm{Tr}(x_a)\ne0,\nonumber\\
\end{array}
\right.
\end{eqnarray*}
\end{lemma}

We will provide the proof of Theorem \ref{weight1} according to two cases.
\begin{enumerate}
  \item If $2\nmid e$, $p\mid e$ and $p\equiv1\bmod4$, with Lemma \ref{w1}, we have
\begin{eqnarray*}
wt_{i1}&=&\frac{(p-1)^2}{2}p^{e-2},\\
wt_{i2}&=&\frac{(p-1)^2}{2}p^{e-2}+(-1)^i\frac{p-1}{2}p^{\frac{e-1}{2}},\\
wt_{i3}&=&\frac{(p-1)^2}{2}p^{e-2}+(-1)^i\frac{p-1}{2}p^{\frac{e-1}{2}}+(-1)^i\frac{p-1}{2}p^{\frac{e-3}{2}},\\
wt_{i4}&=&\frac{(p-1)^2}{2}p^{e-2}+(-1)^i\frac{p-1}{2}p^{\frac{e-1}{2}}-(-1)^i\frac{p-1}{2}p^{\frac{e-3}{2}},\\
wt_{i5}&=&\frac{(p-1)^2}{2}p^{e-2}+(-1)^i\frac{p-1}{2}p^{\frac{e-1}{2}}-(-1)^i\frac{p+1}{2}p^{\frac{e-3}{2}},
\end{eqnarray*}
and
\begin{eqnarray*}
A_{wt_{i1}}&=&N(0,0)-1,\\
A_{wt_{i2}}&=&N(0,\overline{0}),\\
A_{wt_{i3}}&=&M_{(i,0)},\\
A_{wt_{i4}}&=&M_{(1-i,0)}+M_{(1-i,\overline{0})},\\
A_{wt_{i5}}&=&M_{(i,\overline{0})},
\end{eqnarray*}
Applying into the results of Lemma \ref{F1} and Lemma \ref{F2}, we can get the first part of Theorem \ref{weight1}.
  \item If $2\nmid e$, $p\mid e$ and $p\equiv3\bmod4$, with Lemma \ref{w1}, we have
\begin{eqnarray*}
wt_{i1}&=&\frac{(p-1)^2}{2}p^{e-2},\\
wt_{i2}&=&\frac{(p-1)^2}{2}p^{e-2}+(-1)^{i+\frac{e+3}{2}}\frac{p-1}{2}p^{\frac{e-1}{2}},\\
wt_{i3}&=&\frac{(p-1)^2}{2}p^{e-2}+(-1)^{i+\frac{e+3}{2}}\frac{p-1}{2}p^{\frac{e-1}{2}}+(-1)^{i+\frac{e+1}{2}}\frac{p-1}{2}p^{\frac{e-3}{2}},\\
wt_{i4}&=&\frac{(p-1)^2}{2}p^{e-2}+(-1)^{i+\frac{e+3}{2}}\frac{p-1}{2}p^{\frac{e-1}{2}}-(-1)^{i+\frac{e+1}{2}}\frac{p-1}{2}p^{\frac{e-3}{2}},\\
wt_{i5}&=&\frac{(p-1)^2}{2}p^{e-2}+(-1)^{i+\frac{e+3}{2}}\frac{p-1}{2}p^{\frac{e-1}{2}}+(-1)^{i+\frac{e+1}{2}}\frac{p+1}{2}p^{\frac{e-3}{2}},
\end{eqnarray*}
and
\begin{eqnarray*}
A_{wt_{i1}}&=&N(0,0)-1,\\
A_{wt_{i2}}&=&N(0,\overline{0}),\\
A_{wt_{i3}}&=&M_{(1-i,0)}+M_{(1-i,\overline{0})},\\
A_{wt_{i4}}&=&M_{(i,0)},\\
A_{wt_{i5}}&=&M_{(i,\overline{0})},
\end{eqnarray*}
Applying into the results of Lemma \ref{F1} and Lemma \ref{F2}, we can get the second part of Theorem \ref{weight1}.
\end{enumerate}

Thus we can get the desired conclusions presented in Theorem \ref{weight1}, and complete the proof.\hfill $\square$

\textbf{Proof of Theorem \ref{weight4}:} Let $2\mid e$, $e\equiv2\bmod4$, and $p\mid e$. With Lemma \ref{length}, for $i=0,1$, we have
\begin{equation*}
n_i=\frac{p-1}{2}p^{e-1}+\frac{p-1}{2}p^{\frac{e}{2}-1}.
\end{equation*}
Applying Lemma \ref{first} and Lemma \ref{second} into Eq.(\ref{5.2.7}), we can get the following lemma directly.
\begin{lemma}\label{w2}
Let $2\mid e$, $e\equiv2\bmod4$, and $p\mid e$, for $i=0,1$, we have
\begin{eqnarray*}
wt_{i}(\mathbf{c}(a))&=&\left\{
\begin{array}{ll}
\frac{(p-1)^2}{2}p^{e-2}, & \mathrm{Tr}(x_a^{p+1})=0, \mathrm{Tr}(x_a)=0,\nonumber\\
\frac{(p-1)^2}{2}p^{e-2}+\frac{p-1}{2}p^{\frac{e}{2}-1}, & \mathrm{Tr}(x_a^{p+1})=0, \mathrm{Tr}(x_a)\ne0,\nonumber\\
\frac{(p-1)^2}{2}p^{e-2}+\frac{1+(-1)^i\eta(-\mathrm{Tr}(x_a^{p+1}))}{2}(p-1)p^{\frac{e}{2}-1},& \mathrm{Tr}(x_a^{p+1})\ne0, \mathrm{Tr}(x_a)=0,\nonumber\\
\frac{(p-1)^2}{2}p^{e-2}+\frac{p-1}{2}p^{\frac{e}{2}-1}-\frac{1+(-1)^i\eta(-\mathrm{Tr}(x_a^{p+1}))}{2}p^{\frac{e}{2}-1}, & \mathrm{Tr}(x_a^{p+1})\ne0, \mathrm{Tr}(x_a)\ne0,\nonumber\\
\end{array}
\right.
\end{eqnarray*}
\end{lemma}
With Lemma \ref{w2}, we can proof Theorem \ref{weight4}.
\begin{eqnarray*}
wt_{i1}&=&\frac{(p-1)^2}{2}p^{e-2},\\
wt_{i2}&=&\frac{(p-1)^2}{2}p^{e-2}+\frac{p-1}{2}p^{\frac{e}{2}-1},\\
wt_{i3}&=&\frac{(p-1)^2}{2}p^{e-2}+(p-1)p^{\frac{e}{2}-1},\\
wt_{i4}&=&\frac{(p-1)^2}{2}p^{e-2}+\frac{p-3}{2}p^{\frac{e}{2}-1},
\end{eqnarray*}
and
\begin{eqnarray*}
A_{wt_{i1}}&=&N(0,0)-1+M_{(1-i+\frac{p-1}{2},0)},\\
A_{wt_{i2}}&=&N(0,\overline{0})+M_{(1-i+\frac{p-1}{2},\overline{0})},\\
A_{wt_{i3}}&=&M_{(i+\frac{p-1}{2},0)},\\
A_{wt_{i4}}&=&M_{(i+\frac{p-1}{2},\overline{0})},
\end{eqnarray*}
Applying into the results of Lemma \ref{F1} and Lemma \ref{F2}, we can get the second part of Theorem \ref{weight4}.\hfill $\square$

\textbf{Proof of Theorem \ref{weight8}:} Let $2\mid e$, $e\equiv0\bmod4$, and $\big(\frac{e}{p}\big)=(-1)^i$. With Lemma \ref{first}, for $i=0,1$, we have
\begin{equation*}
n_i=\frac{p-1}{2}p^{e-1}-\frac{1+\eta(-1)p}{2}p^{\frac{e}{2}}.
\end{equation*}
Applying Lemma \ref{first} and Lemma \ref{second} into Eq.(\ref{5.2.7}), we can get the following lemma directly.
\begin{lemma}\label{w3}
Let $2\mid e$, $e\equiv0\bmod4$, and $\big(\frac{e}{p}\big)=(-1)^i$, for $i=0,1$, we have
\begin{enumerate}
\item if there are no solutions of the equation $x^{p^2}+x=-a^p$ in $\mathbb{F}_q$, then
\begin{equation*}
wt_{i}(\mathbf{c}(a))=\frac{(p-1)^2}{2}p^{e-2}-\frac{1+\eta(-1)p}{2}(p-1)p^{\frac{e}{2}-1},
\end{equation*}
\item if $x_a$ is one solution of the equation $x^{p^2}+x=-a^p$ in $\mathbb{F}_q$, then
\begin{eqnarray*}
wt_{i}(\mathbf{c}(a))&=&\left\{
\begin{array}{ll}
\frac{(p-1)^2}{2}p^{e-2}, \qquad\qquad\qquad\qquad\qquad\qquad\quad \mathrm{Tr}(x_a^{p+1})=0, \mathrm{Tr}(x_a)=0,\nonumber\\
\frac{(p-1)^2}{2}p^{e-2}-\frac{1+\eta(-1)p}{2}p^{\frac{e}{2}}, ~\qquad\qquad\qquad\quad \mathrm{Tr}(x_a^{p+1})=0, \mathrm{Tr}(x_a)\ne0,\nonumber\\
\frac{(p-1)^2}{2}p^{e-2}+\frac{1-\eta(-1)p}{2}p^{\frac{e}{2}}-(1+(-1)^i\eta(-\mathrm{Tr}(x_a^{p+1})))p^{\frac{e}{2}},\nonumber\\ ~~\qquad\qquad\qquad\qquad\qquad\qquad\qquad\qquad\quad \mathrm{Tr}(x_a^{p+1})\ne0, \mathrm{Tr}(x_a)=0,\nonumber\\
\frac{(p-1)^2}{2}p^{e-2}-\frac{1+\eta(-1)}{2}p^{\frac{e}{2}},\nonumber\\
\qquad\qquad\qquad\qquad\quad \mathrm{Tr}(x_a^{p+1})\ne0, \mathrm{Tr}(x_a)\ne0, e\mathrm{Tr}(x_a^{p+1})=\mathrm{Tr}(x_a)^2,\nonumber\\
\frac{(p-1)^2}{2}p^{e-2}-\frac{1+\eta(-1)p}{2}p^{\frac{e}{2}}
-\frac{(-1)^i\eta(-\mathrm{Tr}(x_a^{p+1}))+\eta(\mathrm{Tr}(x_a)^2-e\mathrm{Tr}(x_a^{p+1}))}{2}p^{\frac{e}{2}}\nonumber\\
\qquad\qquad\qquad\qquad\quad \mathrm{Tr}(x_a^{p+1})\ne0, \mathrm{Tr}(x_a)\ne0, e\mathrm{Tr}(x_a^{p+1})\ne\mathrm{Tr}(x_a)^2,\nonumber\\
\end{array}
\right.
\end{eqnarray*}
\end{enumerate}
\end{lemma}
We will give the proof process of Theorem \ref{weight8} according to the following two cases. With Lemma \ref{w3},
\begin{enumerate}
  \item if $2\mid e$, $e\equiv0\bmod4$, $\big(\frac{e}{p}\big)=(-1)^i$, and $p\equiv1\bmod4$, we have
      \begin{eqnarray*}
wt_{i1}&=&\frac{(p-1)^2}{2}p^{e-2}-\frac{p^2-1}{2}p^{\frac{e}{2}-1},\\
wt_{i2}&=&\frac{(p-1)^2}{2}p^{e-2},\\
wt_{i3}&=&\frac{(p-1)^2}{2}p^{e-2}-\frac{p+1}{2}p^{\frac{e}{2}},\\
wt_{i4}&=&\frac{(p-1)^2}{2}p^{e-2}-\frac{p+3}{2}p^{\frac{e}{2}},\\
wt_{i5}&=&\frac{(p-1)^2}{2}p^{e-2}-\frac{p-1}{2}p^{\frac{e}{2}},\\
wt_{i6}&=&\frac{(p-1)^2}{2}p^{e-2}-p^{\frac{e}{2}},
\end{eqnarray*}
and
\begin{eqnarray*}
A_{wt_{i1}}&=&p^e-p^{e-2},\\
A_{wt_{i2}}&=&p^{-2}N(0,0)-1,\\
A_{wt_{i3}}&=&p^{-2}N(0,\overline{0})+p^{-2}N(i, \overline{0}, \overline{e}, 1-i)+p^{-2}N(1-i, \overline{0}, \overline{e}, 1-i),\\
A_{wt_{i4}}&=&p^{-2}M_{(i,0)}+p^{-2}N(i,\overline{0}, \overline{e}, i),\\
A_{wt_{i5}}&=&p^{-2}M_{(1-i,0)}+p^{-2}N(1-i,\overline{0}, \overline{e}, i),\\
A_{wt_{i6}}&=&p^{-2}N(\overline{0}, \overline{0}, e),
\end{eqnarray*}
Applying into the results of Lemma \ref{F1}, Lemma \ref{F2}, Lemma \ref{F3} and Lemma \ref{F6}, we can get the first part of Theorem \ref{weight8}.
  \item if $2\mid e$, $e\equiv0\bmod4$, $\big(\frac{e}{p}\big)=(-1)^i$, and $p\equiv3\bmod4$, we have
      \begin{eqnarray*}
wt_{i1}&=&\frac{(p-1)^2}{2}p^{e-2}+\frac{(p-1)^2}{2}p^{\frac{e}{2}-1},\\
wt_{i2}&=&\frac{(p-1)^2}{2}p^{e-2},\\
wt_{i3}&=&\frac{(p-1)^2}{2}p^{e-2}+\frac{p-1}{2}p^{\frac{e}{2}},\\
wt_{i4}&=&\frac{(p-1)^2}{2}p^{e-2}+\frac{p-3}{2}p^{\frac{e}{2}},\\
wt_{i5}&=&\frac{(p-1)^2}{2}p^{e-2}+\frac{p+1}{2}p^{\frac{e}{2}},
\end{eqnarray*}
and
\begin{eqnarray*}
A_{wt_{i1}}&=&p^e-p^{e-2},\\
A_{wt_{i2}}&=&p^{-2}N(0,0)-1+p^{-2}N(\overline{0}, \overline{0}, e),\\
A_{wt_{i3}}&=&p^{-2}N(0,\overline{0})+p^{-2}N(i, \overline{0}, \overline{e}, 1-i)+p^{-2}N(1-i, \overline{0}, \overline{e}, 1-i),\\
A_{wt_{i4}}&=&p^{-2}M_{(1-i,0)}+p^{-2}N(1-i,\overline{0}, \overline{e}, i),\\
A_{wt_{i5}}&=&p^{-2}M_{(i,0)}+p^{-2}N(i,\overline{0}, \overline{e}, i),
\end{eqnarray*}
Applying into the results of Lemma \ref{F1}, Lemma \ref{F2}, Lemma \ref{F3} and Lemma \ref{F6}, we can get the second part of Theorem \ref{weight8}.\hfill$\Box$
\end{enumerate}

The proof process of other theorems is similar to the proof of the above three theorems, and will not be repeated here.

\textbf{Remark 1:} In the case of $e\equiv0\bmod4$, if there are no solutions of the equation $x^{p^2}+x=-a^p$ in $\mathbb{F}_q$, with Lemma \ref{weil sums4}, the number of $a$ is $p^e-p^{e-2}$.

\textbf{Remark 2:} In the case of $e\equiv0\bmod4$, if the equation $x^{p^2}+x=-a^p$ is solvable, then it has $p^2$ solutions.

\textbf{Remark 3:} In the nine theorems, there are some special cases, such as the case of $p=3$, in which the codes may have less number of non-zero weights.

\section{Concluding remarks}\label{section-5}\rm
In this paper, inspired by the work in \cite{CCK}, two classes of at most six-weight linear codes were constructed with their weight enumerators settled using Weil sums and Gaussian period. At the same time, some optimal or almost optimal linear code was found.

Let $wt_{\min}$ and $wt_{\max}$ denote the minimum and maximum non-zero weight of a linear code $\mathcal{C}$, respectively. Obviously, if $e\geq8$,
$$\frac{wt_{\min }}{wt_{\max }} > \frac{p - 1}{p}.$$
As stated in \cite{YJD}, the codes $\mathcal{C}_D$ can be used to construct secret sharing schemes.

\end{document}